\documentclass[12pt,preprint]{aastex}
\usepackage{amsmath}

\begin{document}


\title{Tidal disruption and ignition of white dwarfs by moderately
  massive black holes} \author{S. Rosswog$^{1}$, E. Ramirez-Ruiz$^{2}$
  and W.R.Hix$^{3}$\\ $^{1}$ School of Engineering and Science, Jacobs
  University Bremen, Campus Ring 1, 28759 Bremen, Germany\\ $^{2}$
  Department of Astronomy and Astrophysics, University of California,
  Santa Cruz, CA 95064\\ $^{3}$Physics Division, Oak Ridge National
  Laboratory, Oak Ridge, TN37831-6374 }

\def\paren#1{\left( #1 \right)}
\def\Mesz{M\'esz\'aros~}
\def\Pacz{Paczy\'nski~}
\def\Kluz{Klu\'zniak~}
\def\p{$e^\pm \;$}
\def\msun{M$_{\odot}$}
\def\Msun{M$_{\odot}$ }
\def\be{\begin{equation}}
\def\ee{\end{equation}}
\def\bi{\begin{itemize}}
\def\ei{\end{itemize}}
\def\bea{\begin{eqnarray}}
\def\eea{\end{eqnarray}}
\def\gcc{gcm$^{-3}$ }
\def\edo{\end{document}}

\begin{abstract}
We present a numerical investigation of  the tidal disruption of white
dwarfs by moderately massive black holes, with particular reference to
the centers of dwarf galaxies and globular clusters.  Special
attention is given to the fate of white dwarfs of all masses that
approach the black hole close enough to be disrupted and severely compressed
to such extent that explosive nuclear burning can be triggered.  
Consistent modeling of the gas dynamics together with
the nuclear reactions allows for a realistic determination of 
the explosive energy release. In the most favorable cases, the nuclear
energy release may be comparable to that of typical type Ia
supernovae. Although the explosion will increase the mass fraction
escaping on hyperbolic orbits, a good fraction of the debris
remains to be swallowed by the hole, causing a bright soft X-ray flare 
lasting for about a year. Such
transient signatures, if detected, would be a compelling testimony for
the presence of a moderately mass black hole (below $10^5 M_\odot$).
\end{abstract}

\maketitle

\section{Introduction}
White dwarfs, the end point of stellar evolution for stars with masses
from about 0.07 to 10 solar masses, are extremely common
\citep{w1990}. And we don't have to look far away, either. There are
several billions of them in the halo of our very own Milky Way galaxy
\citep{r2005}. They are not only observed in isolation but in binary
systems, with normal stellar companions, and, less frequently, with
compact stellar companions. At a distance of tens of kiloparsecs, a
number of globular clusters \citep{br2006,m2008} have a high enough
central density to let white dwarfs interact and collide with other
stars, or, if developed, with a central massive black hole
\citep{ge2002,ge2003,geb2002,geb2005}.  And at a distance of less than
ten kiloparsecs, our galactic nucleus, a central massive black hole of
a few million solar masses \citep{gh1998}, is surrounded by swarms of
all kind of stars, some of them white dwarfs and all of them prone to
collisions. 
In such dense environments, stars of all varieties
can exchange mass, disrupt each other or merge, and their merger
products can get involved in similar interactions. 
Because of this reason, the innermost $\sim 0.1$ pc of our Galaxy can 
be considered an efficient stellar collider \citep{alexander05}.
Stellar dynamics is
perfectly adequate in modeling the motions of stars as point masses
moving under the influence of gravity, even in dense stellar systems,
unless individual stars closely approach each other or get so close to
the central black hole that they become vulnerable to tidal
distortions. Once this happens, the internal structure of 
the stars becomes important for the further dynamical evolution. 
This paper explores the observational manifestations of such phenomena, 
with particular reference to white dwarfs.

A white dwarf interacting with a neutron star or a black hole cannot
be treated as a point mass if it gets so close to the compact remnant
that it becomes tidally deformed.  The tidal interaction 
between a star and a black hole is mainly characterized by three
length scales: the stellar radius, $R_{\rm wd}$, the gravitational
radius of the black hole, $R_{\rm g}= 2GM_{\rm BH}/c^2\simeq 3 \times
10^{11} M_{\rm BH,6}$ cm, where $M_{\rm BH,6}$ denotes the hole's mass in units of $10^6 M_\sun$, and the tidal radius, $R_{\tau}$.
The tidal radius, defined as the distance within which a stars gets 
disrupted, obviously depends on the type of star
being considered. For a white dwarf it is roughly given by
\be 
R_\tau
\simeq 1.2 \times 10^{11} M_{\rm BH,6}^{1/3} \left({R_{\rm wd} \over
  10^{9}{\rm cm}}\right)\left({M_{\rm wd} \over 0.6
  M_\sun}\right)^{-1/3} {\rm cm}. 
\ee
Essentially, $R_\tau$ is the distance from the hole at which 
$M_{\rm BH}/R_\tau^3$ equals the mean density of the passing star. 
$R_\tau$ is, in order of magnitude, the
same as the Roche radius, which is a precisely defined
quantity, but only strictly applicable to a star in a circular orbit
with synchronized spin. The strength of the tidal encounter is
measured by the dimensionless parameter 
\be \eta=\left({M_{\rm wd} \over
  M_{\rm BH}}{R_{\rm p}{^3} \over R_{\rm wd}{^3}}\right)^{1/2}, 
\ee 
which is simply the square root of the ratio of the surface 
gravity of the star to the tidal gravity at the surface when the 
star is at pericenter distance $R_p$. Most literature on tidal 
disruption uses $\beta=R_\tau/R_p=\eta^{-2/3}$ to measure the 
encounter strength. When $\eta \leq 1$,
the star is disrupted in a single flyby. The energy required to tear
the star apart (that is, the star's self-binding energy) is supplied
at the expense of the orbital energy, which at $R_\tau$ is larger by
$(M_{\rm BH}/M_{\rm wd})^{2/3}$.

\begin{figure}
\epsscale{1.0}
\plotone{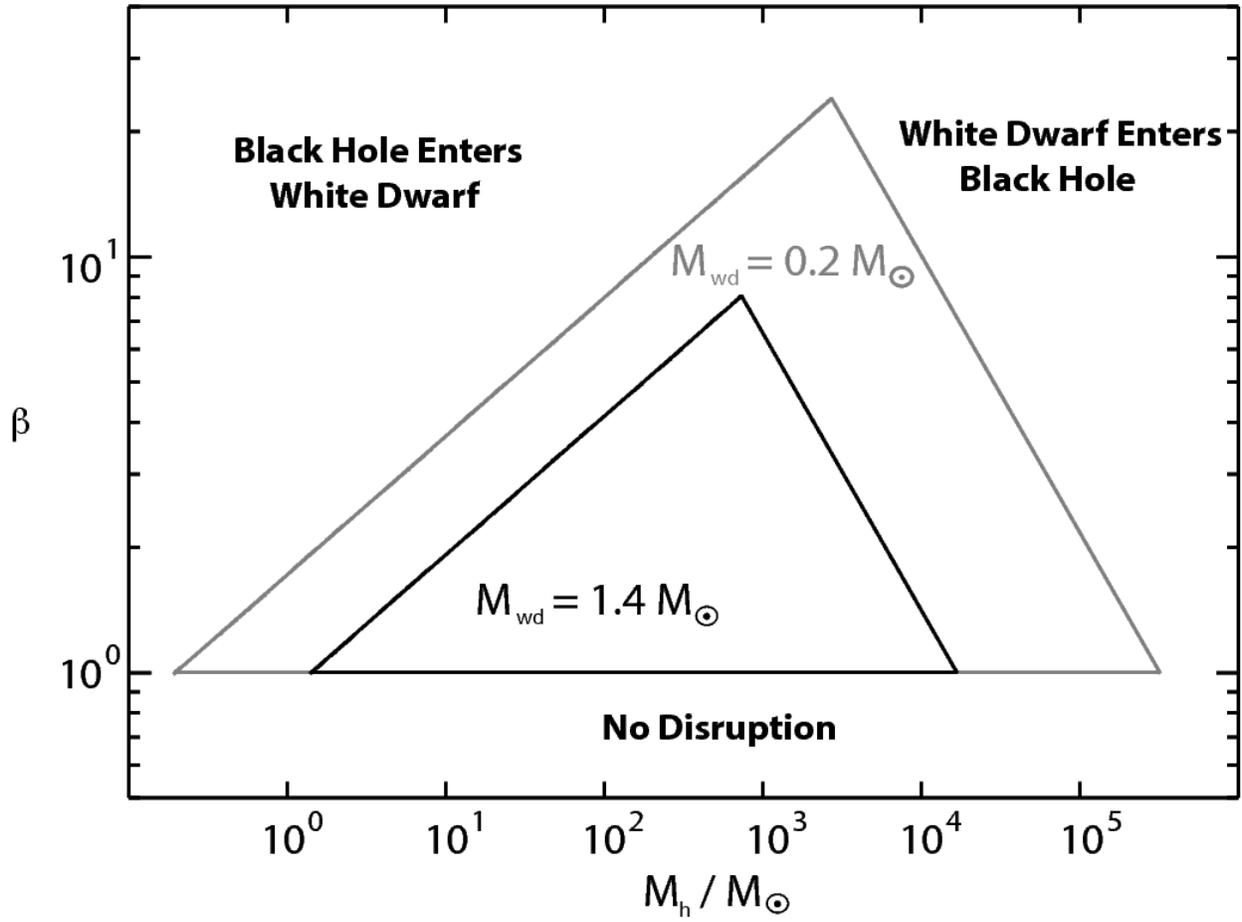}
\caption{This diagrams depicts the relevant domain where the black
  hole's tidal effect can be disruptive to a white dwarf star. The
  penetration factor $\beta$ is plotted as a function of the black
  hole mass $M_{\rm BH}$. Note that a black hole of $\gg 10^5 M_\odot$
  can swallow all white dwarfs without first disrupting them (adapted
  from Luminet \& Pichon 1989).}
\label{fig1}
\end{figure}

For white dwarfs, the ratio between the tidal disruption radius and
the gravitational radius is 
\be 
\beta_{\rm g}=\left({R_\tau \over
  R_{\rm g}}\right)  \simeq 0.4 M_{\rm BH,6}^{-2/3} \left({R_{\rm wd}
  \over 10^{9}{\rm cm}}\right)\left({M_{\rm wd} \over 0.6
  M_\sun}\right)^{-1/3}.  
\ee 
Note that this is indeed inside the
gravitational radius for black hole masses exceeding 
\be M_{\rm BH,
  lim} \simeq 2.5 \times 10^{5} \left({R_{\rm wd} \over 10^{9}{\rm
    cm}}\right)^{3/2}\left({M_{\rm wd} \over 0.6
  M_\sun}\right)^{-1/2}\; M_\sun.  
\ee
 For this reason, white dwarfs
only experience disruptive physical conditions when
approaching a moderately massive black hole \citep{lu1989}. If $M_{\rm
  BH} \ll M_{\rm BH,lim}$ it is however, sufficiently far outside $R_{\rm
  g}$ that the disruption can be approximated as a Newtonian process,
and it makes little difference whether the hole is described by a
Schwarzschild or a Kerr metric. When the central hole has a mass
$M_{\rm BH} \gg M_{\rm wd}$, the size of the star remains smaller than
$R_\tau$.  A white dwarf cannot thus be disrupted (Figure~\ref{fig1})
without entering the strong relativistic domain if \be \beta > \min
\left[\left({M_{\rm BH} \over M_{\rm wd}}\right)^{1/3}, \left({M_{\rm
      BH, lim}\over M_{\rm BH}}\right)^{2/3}\right].  \ee The type of
the black hole (Schwarzchild or Kerr?)  then has an important
quantitative effect, as does (for a Kerr hole) the orientation of the
stellar orbit relative to the spin axis.

Much of our effort in this paper is therefore dedicated to investigate
the disruption of white dwarfs by moderately massive black holes, with
particular reference to the centers of dwarf galaxies
\citep[e.g.][]{ma2008}, globular clusters \citep[e.g.][]{po2004,
  ba2004} and the intermediate-mass black hole candidates in active 
galactic nuclei \citep{greene04}.  Special attention is given to the 
fate of white dwarfs of
all masses that approach a black hole close enough to be 
disrupted and severely compressed to such extent that explosive
nuclear burning can be triggered. Tidal disruption of white dwarfs has
previously been difficult to model due to its three-dimensional
nature and the delicate interplay between gravity, gas 
dynamics and nuclear reactions. Crucial aspects of the physics 
of white dwarf disruption and ignition were first understood by 
\citet{lu1989}, mainly based on the affine model earlier 
introduced by \citet{cl1983}. \citet{lu1989} suggested that
for very close encounters ($\eta \leq 1$), relativistic effects and
compression into the orbital plane could trigger significant nuclear
energy release in the core (some authors, particularly Wilson \&
Mathews 2004 and Dearborn et al. 2005, believe such considerations 
could also apply to weakly
disrupted encounters for which general relativistic effects can
overwhelm the stabilizing tidal forces and still cause relativistic
compression). The affine model of Carter and Luminet approximates
the white dwarf as a time-dependent ellipsoid, with a fixed density profile. 
Such a model is of limited utility if the true matter distribution deviates 
from the assumed shape and density profile. To study the tidal disruption 
problem, we use
a three dimensional smoothed particle hydrodynamics method (SPH) to
solve the equations of hydrodynamics. This method makes no restrictive
assumptions on the hydrodynamic degrees of freedom and allows us to
study the three-dimensional encounter with full generality. Due to its
Lagrangian nature SPH is perfectly suited to follow tidal disruption
processes during which the corresponding geometry, densities and time
scales are changing violently \citep{ro2008,rrh2008}. What is more,
our consistent modeling of gas dynamics together with 
nuclear energy generation allows for a realistic determination
of the explosive energy release.

In this paper, a comprehensive approach is adopted: we follow the
tidal disruption and compression leading up to the ignition of the 
white dwarf (when triggered); the accompanying nucleosynthesis and
expansion of the debris; and the subsequent accretion of the bound
debris. Some of the questions at the forefront of attention include
the effects of initial composition and mass of the white dwarf, 
the passage distance, and the black hole's mass on ignition
and disruption, and the degree to which the post-disruption dynamics
is modified by nuclear energy release. We address all of these issues
here.\\
A detailed description of the numerical methods and the initial models
is given in \S \ref{nm}. Detailed hydrodynamic simulations of the 
disruption of white dwarfs of various masses, initial composition and 
passage distances are presented in \S\S \ref{wd}, \ref{bh} and \ref{var},
the role of the mass of the black hole in shaping the evolution and 
ignition of the disrupted white dwarf is discussed in \S \ref{bh}.  
The resulting gravitational wave signals are shown in \S \ref{gw}. 
Discussion and conclusions are presented in \S \ref{dis}.

\section{Numerical Methods and Initial Model}\label{nm}
The observational consequences of stellar disruption depend on what
happens to the debris \citep{re1988,ev1989,ko2004}. To quantify this,
we have performed detailed three-dimensional, hydrodynamical
calculations. The gas dynamics is coupled with a nuclear network to
explore the effects of nuclear energy generation during the strong
compression phase. The reader is referred to \citet{ro2008} for a
complementary description of the numerical methods employed in
calculating the disruptive event.

\subsection{Hydrodynamics}
The SPH-formulation used in this study follows closely the one
described in \citet{be1990}, a brief derivation of the equations
can be found in \citet{rosswog08c}. Forces from self-gravity are calculated
using a binary tree \citep{bet1990} while the gravitational forces
from the central black hole are calculated using a Paczy\'nski-Wiita
pseudo potential \citep{pa1980} with a polynomial extension 
\citep{rosswog05} to avoid the singularity at the Schwarzschild 
radius.
We have taken particular care to avoid artifacts from the use of 
artificial viscosity (AV). The
so-called Balsara-switch \citep{ba1995} is implemented to avoid
spurious shear forces. More importantly, we use time dependent 
viscosity parameters \citep{mo1997}, so that far from shocks
artificial viscosity is essentially absent, but near a shock front the
associated viscous parameters rise to values that are able to avoid post-shock
oscillations (see Fig. 1 of \citep[][]{ro2008} for an illustration). 
Once the shocks has passed, the parameters decay again. Details of the 
AV implementation and tests can be found in \citep{ro2000}.

A MacCormack predictor-corrector method is used to evolve the fluid in
time. The method is implemented with individual time steps, i.e. each
particle $i$ is advanced on its own time step $dt_i= 2^{n_i} \times
dt_{\rm min}$, where $dt_{\rm min}$ is the smallest step of all the
particles. $n_i$ is chosen to be the largest integer satisfying the
condition $dt_i < dt_{i,{\rm des}}$, where $dt_{i, {\rm des}}$ is the
desired time step for particle $i$. With this time marching
implementation the total energy is conserved to better than $4\times
10^{-3}$ and the total angular momentum to better than $2\times
10^{-4}$. Note that this could, in principle, be improved even further
by taking into account the so-called ``grad-h''-terms \citep{springel02,mo2002}
and extra-terms arising from adapting gravitational smoothing terms 
\citep{price07a}. A comparison between the formulation used here
and the one using  ``grad-h''-terms can be found in \citet{rosswog07}.

\subsection{Equation of State}
We use the HELMHOLTZ equation of state (EOS), developed by the 
Center for Astrophysical Thermonuclear Flashes at the University of Chicago. 
The EOS allows to freely specify the chemical composition of
the gas and it can be coupled to a nuclear reaction network. 
The electron-/positron equation of state is calculated without
approximation. In other words, it makes no assumption about the degree
of degeneracy or relativity and the exact expressions are integrated
numerically. The nuclei in the gas are treated as a Maxwell-Boltzmann
gas, the photons as blackbody radiation. The EOS is used in tabular
form with densities in the range $10^{-10} \le \rho Y_e \le 10^{11}$ g
cm$^{-3}$ and temperatures varying between $10^4$ and $10^{11}$ K. A
sophisticated, biquintic Hermite polynomial interpolation is used to
enforce the thermodynamic consistency (i.e. the Maxwell-relations) at
interpolated values \citep{timmes00}.

\subsection{Nuclear Burning}
To account for the feedback onto the fluid from nuclear transmutations
we use a minimal nuclear reaction network developed by
\citet{hi1998}. It couples a conventional $\alpha$-network stretching
from He to Si with a quasi-equilibrium-reduced
$\alpha$-network. The QSE-reduced network  
neglects reactions within small equilibrium groups 
that form at temperatures above $3.5 \times 10^9$ K to reduce the 
number of abundance variables needed. Although a set of only 
seven nuclear species is used, this network reproduces all burning 
stages from He-burning to NSE accurately. For details and tests 
we refer to \citet{hi1998}.

The network is coupled to the hydrodynamics in an operator splitting
fashion, i.e. hydrodynamics and nuclear burning -which may in extreme
cases require vastly different time steps- are integrated
separately. In a first step, the hydrodynamic equations 
are integrated with the above described predictor-corrector 
scheme to obtain new quantities at time step $t^{n+1}$.
In this step we ignore the nuclear source term in the energy 
equation, the new value for the specific energy of particle $a$ 
is denoted by $\tilde{u}_a^{n+1}$. The specific energy has to be 
corrected for the energy release
that occurred between $t^{n}$ and $t^{n+1}$:
\begin{eqnarray}
\epsilon_{a,n\rightarrow n+1}&=& - N_{\rm A} \sum_j m_j c^2
\int_{t^{n}}^{t^{n+1}} \frac{dY_{j,a}}{dt} (\rho_a(t),T_a(t),Y_{k,a}(t)) \;  dt\\
&=& - N_A \sum_{j=1} m_j c^2 (Y_{j,a}^{n+1}-Y_{j,a}^n),
\end{eqnarray}
where $N_{\rm A}$ is Avogadro's constant, $m_j c^2$ is the mass energy 
of nucleus $j$ and $Y_{j,a}$ is the abundance of nucleus $j$ in 
particle $a$. We use $\rho_a(t) \approx \rho_a(t^n) + \frac{t-t^n}{t^{n+1}-t^n}
\{\rho_a(t^{n+1})-\rho_a(t^n)\}$ and $T_a(t)\approx \frac{t-t^n}{t^{n+1}-t^n}
\{\tilde{T}_a(t^{n+1})-T_a(t^n)\}$, where $\tilde{T}_a(t^{n+1})$ is the temperature
derived from $\tilde{u}_a^{n+1}$, to integrate the abundances $Y_{j,a}$ via the 
implicit backward Euler method (the network integration is described in 
detail in \citep[][]{hi1998})\footnote{Note that the temperature 
evolution along a hydrodynamical time step is different from 
the description in \citet{ro2008}. In practice, we only see tiny differences 
between both approaches.}. 
The final value for the specific energy at 
time $t^{n+1}$ is given by
\be
u_a^{n+1}= \tilde{u}_a^{n+1} + \epsilon_{a,n\rightarrow n+1}.
\ee
The EOS is then called again to make all thermodynamic quantities 
consistent with this new value $u_a^{n+1}$. Once the derivatives have 
been updated, the procedure can be repeated for the next time step.
For details of the time step criteria we refer the reader to \citet{ro2008}.\\
As will be shown below, the peak compression occurs at a spatially fixed point
(see density panels in Fig.~\ref{fig6}). The white dwarf fuel is fed with 
free-fall velocity $v_{\rm ff}= (2 G M_{\rm BH}/R_{\rm p})^{1/2}
= 1.6 \cdot 10^{5} \; {\rm km \; s}^{-1} \; \left(M_{\rm BH}/1000 \;
{\rm M}_{\odot}\right)^{1/2} \left(R_{\rm p}/10^4 \; {\rm km}\right)^{-1/2}$ 
into this compression point. This is many orders of magnitude larger than
typical flame propagation speeds, therefore flame propagation effects 
can be safely neglected for this investigation.

\subsection{Initial Model}
\begin{figure}
\epsscale{1.0}
\plotone{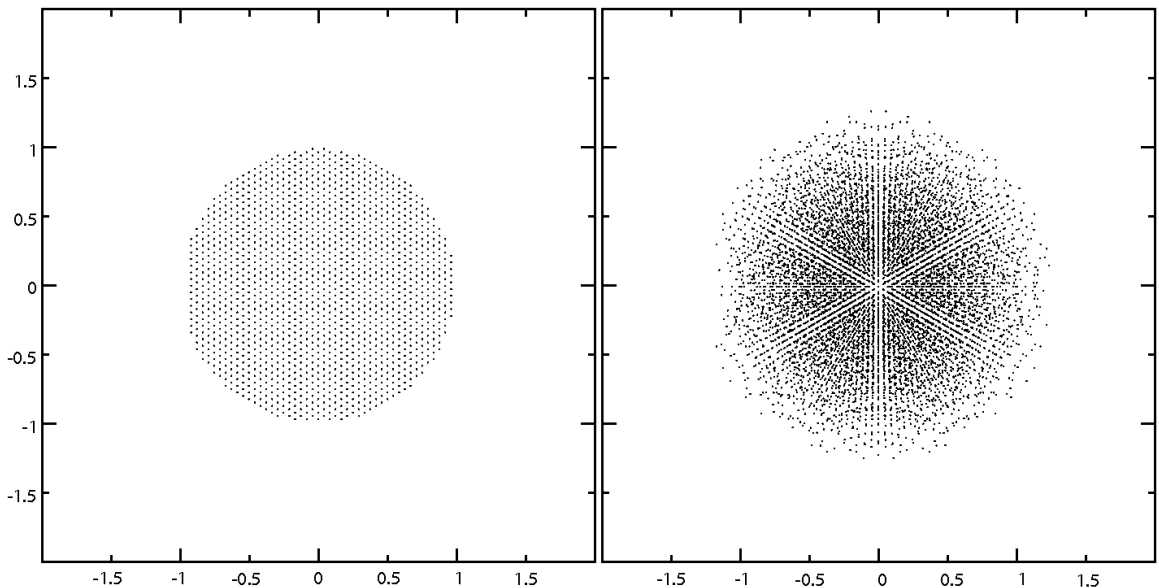}
\caption{Building initial conditions by solving the spherically
  symmetric Lane-Emden equations for a star of given mass and
  composition to find a one dimensional density profile. The 
  particles are then distributed inside a unit sphere according to a
  close-packed prescription.  {\it Left Panel:} Close-packed particle
  distributions in a unit sphere. This distribution is then mapped
  into the volume of the star that is to be constructed. {\it Right
    Panel:} Particle distribution after mapping onto the density
  profile of a 0.2 $M_\odot$ white dwarf. Shown are the projections of
  the particle positions onto the xy-plane. This particle distribution 
  is then finlly 'relaxed' into its true numerical equilibrium.}
\label{fig2}
\end{figure}

It is a vital ingredient for any calculation to start out from initial
conditions that are as accurate as possible. In order to build stars
in hydrostatic equilibrium, it is therefore necessary to find SPH 
particle distributions whose number density reflects the equilibrium mass
density distribution. To this end we solve the spherically symmetric
Lane-Emden equations for a star of given mass and composition to find
a one dimensional density profile.
For simplicity, and to start from a conservatively low 
value, we assume a uniform temperature of  
$T_0= 5\times 10^4$ K. This assumption will be relaxed in one run
(run 10, see Table 1) to test for the sensitivity to this initial
condition. Once the solution to the Lane-Emden equation has been 
obtained, we distribute the desired number of particles,
$N$, inside a unit sphere according to a close-packed configuration. 
Subsequently, we map this distribution into the
volume of the star that is to be constructed. As an example, the
particle distributions before and after mapping of a 0.2 $M_\odot$
white dwarf are shown in Figure~\ref{fig2}.

\begin{figure}
\plotone{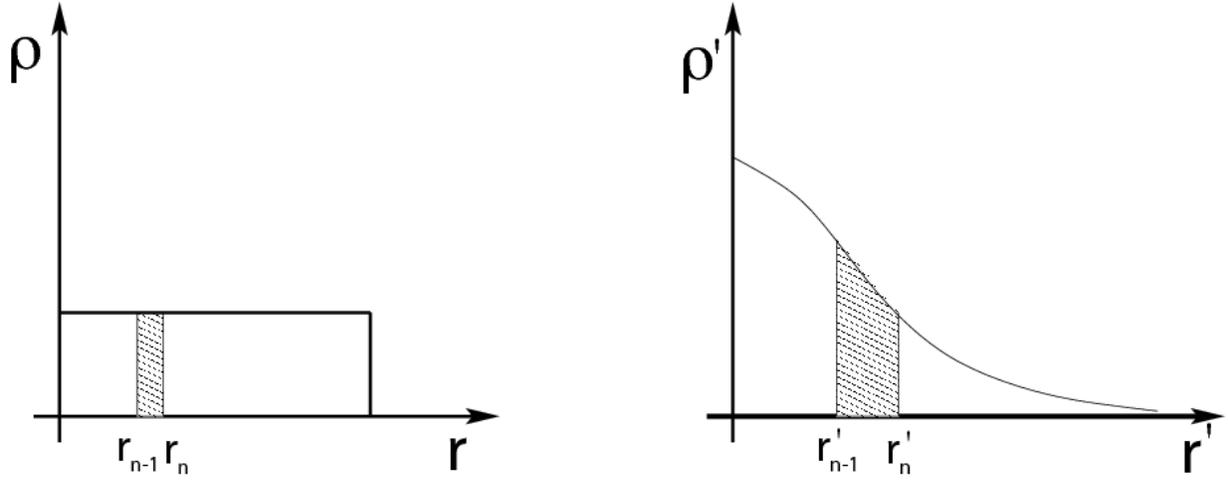}
\caption{Mapping between unit sphere and star. We denote quantities
  referring to the star (unit sphere) with (un-)primed variables.}
\label{fig3}
\end{figure}

The mapping of the initial configuration is done so that the
SPH-particle number density in the star, $n$, yields 
\be
 \rho_{\rm LE}(r)= m \cdot n(r), 
\ee
where $m$ is the mass of each particle and
$\rho_{\rm LE}$ is the density profile obtained by solving the
Lane-Emden equation. We denote quantities referring to the star (unit
sphere) with (un-)primed variables. If the unit sphere has a constant
density $\rho_0$ and contains $N$ particles, each with mass $m= M_{\rm
  wd}/N$, the mass in the shell between $r_{n-1}$ and $r_{n}$ is $M_n
\simeq 4 \pi r_{n-1}^2 \Delta r \rho_0$. As illustrated in
Figure~\ref{fig3}, the image of this shell (in the star) contains
the same mass, but now between $r'_{n-1}$ and $r'_{n}$, so that $M'_n
\simeq 4 \pi r'^2_{n-1} \Delta r'_n \rho_{n-1}$. The condition
$M'_n=M_n$ then yields \be \Delta r'_n= \left(
\frac{r_{n-1}}{r'_{n-1}} \right)^2 \left(\frac{\rho_0}{\rho'_{n-1}}
\right) \Delta r. \ee The resultant configuration is very close to
hydrostatic equilibrium. To find the ``real'' numerical equilibrium
state we relax this configuration with the full hydrodynamics code by
applying an artificial, velocity dependent damping force as, for
example, in \citet{ro2004}.

We use this procedure for all but the heaviest white
dwarfs (1.2 $M_\odot$). Since their adiabatic exponent is already 
approaching the critical 4/3, these stars are  very centrally 
condensed and, as a result, the bulk of the SPH particles resides very 
close to each other in the stellar center. This would, even for a moderate 
resolution of a few hundred thousand particles, result in most of the 
particles running at time steps that are orders of magnitude smaller than 
the dynamical time of the star. To alleviate this computational burden in 
the heaviest white dwarfs we allow for slightly different particle masses 
(about a factor of 10 from the center to the stellar surface).\\
Throughout this simulation set we assume a uniform nuclear
composition accross the white dwarfs. Stars with masses $<$ 0.6 \Msun
are instantiated as pure helium, more massive stars are modeled
as 50\% carbon and  50\% oxygen. Note that helium core burning
may produce composition gradients throughout the stars with oxygen
being more abundant in the stellar core. Since it is mainly the core
that becomes ignited during the compression, see Sect. \ref{wd},
such a gradient may slightly reduce the nuclear energy that can be 
released in principle.\\
Common to all
calculations is the initiation of the calculations with the white
dwarf being placed safely outside $R_{\tau}$. Initial separations 
are at least twice, in most cases three times the tidal radius
to allow the initially spherical star to adjust properly to the
changing tidal potential as it approaches the black hole. All
stars are set onto parabolic orbits so that $R_{\rm p}$ is larger than 
$R_{\rm g}$.\\
The performed runs are summarized in Table \ref{tab:runs}.

\section{Events in the Life  of a Tidally Disrupted White Dwarf}\label{wd}
\begin{figure}
\epsscale{0.8}
\plotone{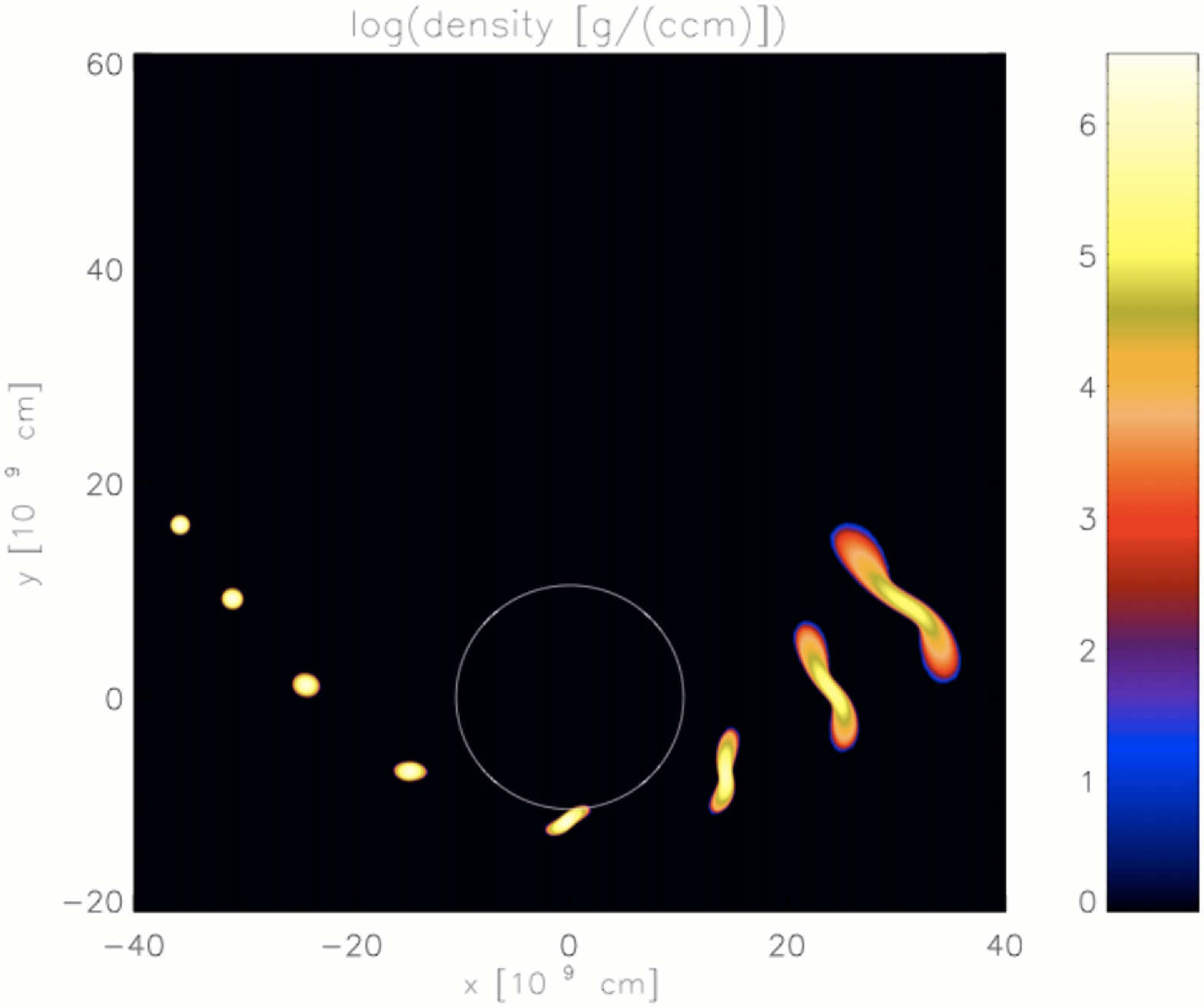}
\epsscale{0.79}
\plotone{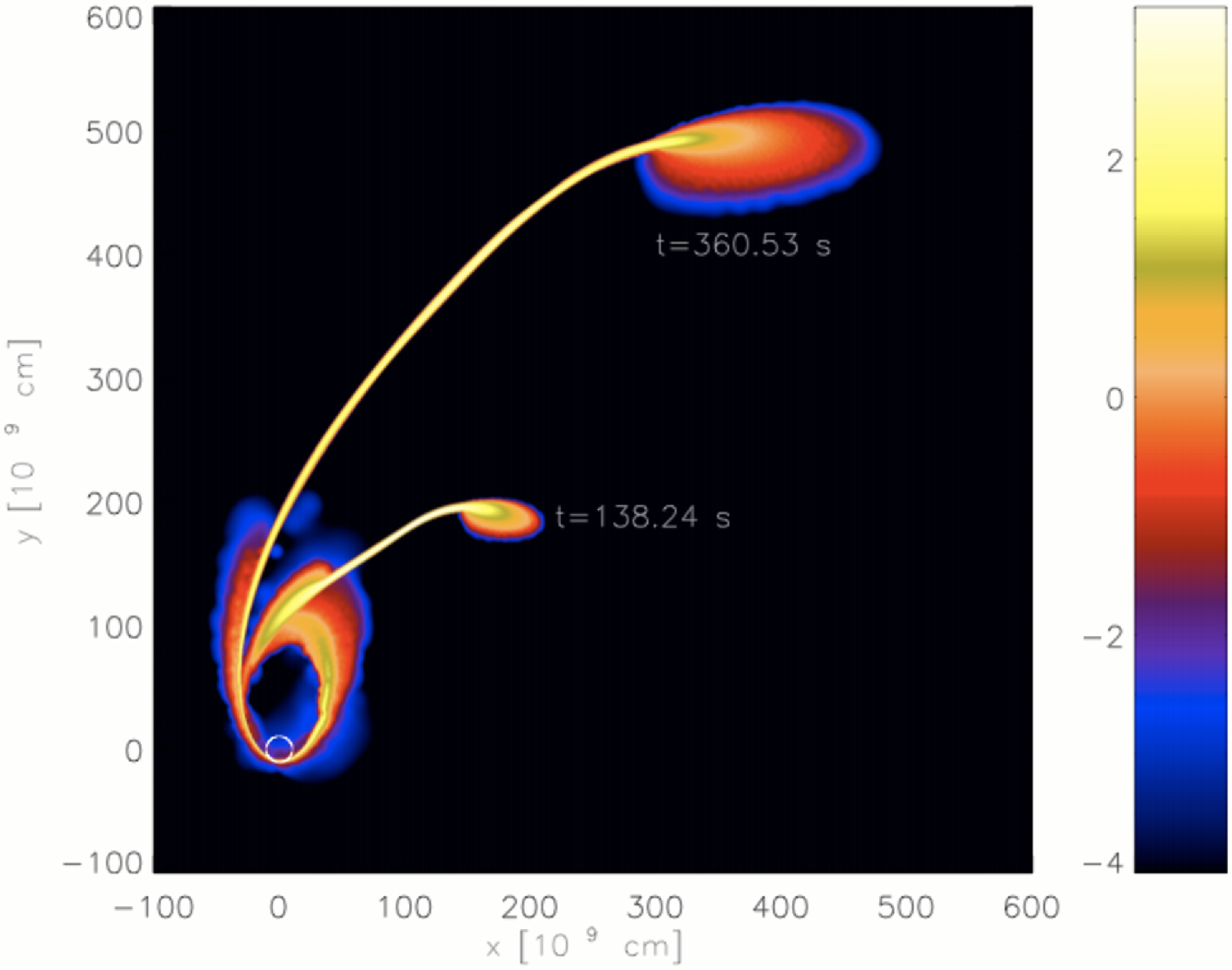}
\caption{Evolution of a 0.6 \Msun white dwarf passing a 1000 \Msun black hole with a penetration factor $\beta$ of only 0.9.
The snapshots show t= 0.34, 3.43, 6.86, 10.29, 13.72, 17.15, 20.58 and 24.01 s after the simulation start in the upper, 
and 138.24 and 380.53 s in the lower panel.}
\label{fig4}
\end{figure}

\begin{figure}
\epsscale{1.0}
\plotone{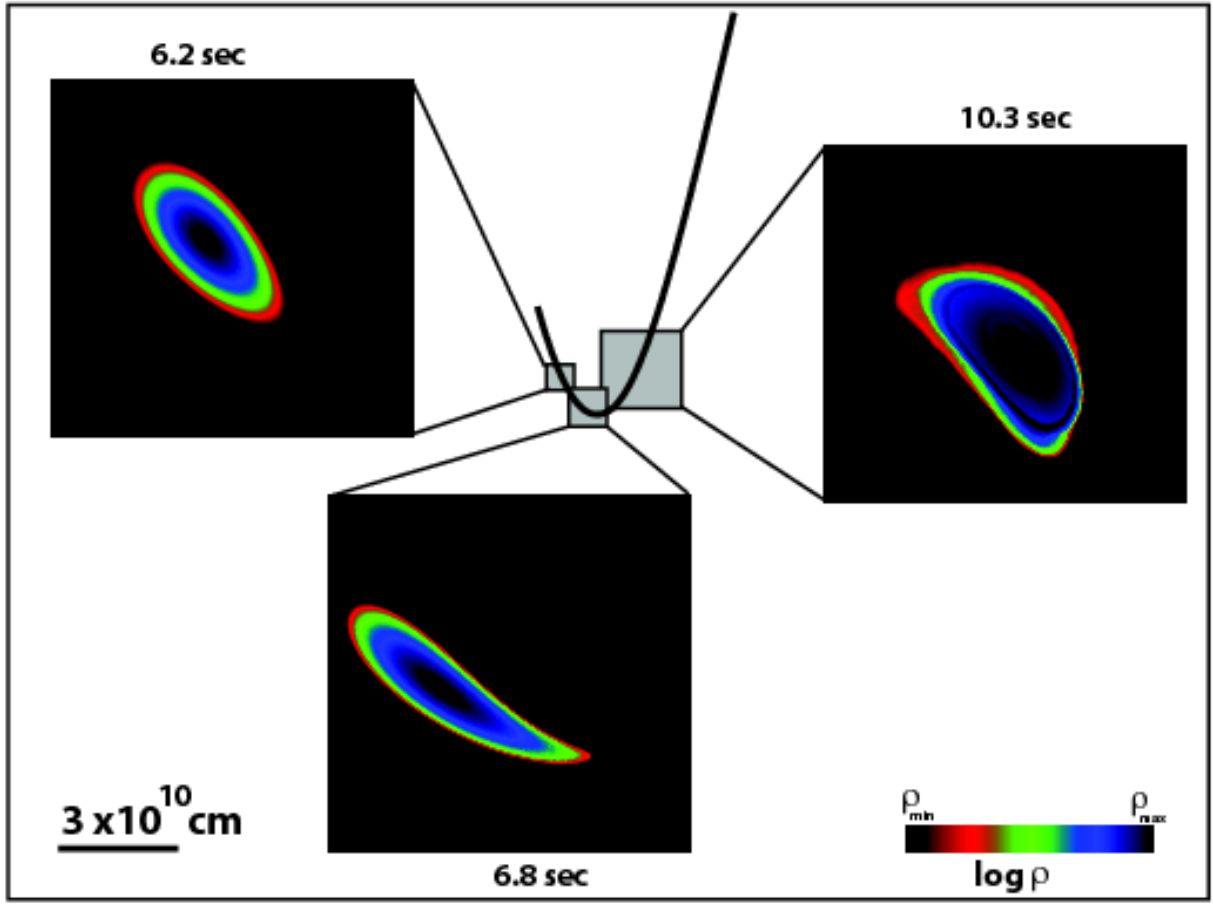}
\caption{A 0.2 $M_\odot$ white dwarf (modeled with more than $4 \times
  10^6$ SPH particles) approaching a $10^3\;M_\odot$ black hole on a
  parabolic orbit with pericenter distance $R_{\rm p} = R_{\tau}/12$ is distorted, spun up during infall and then tidally
  disrupted. Shown are density cuts at various instants along the
  orbital (xy-) plane. Color bar gives the amplitude of $\log
  \rho=[\log \rho_{\rm max}, \log \rho_{\rm min}]$ in cgs units:
  $[4.0,5.3]$ for $t$=6.2 s, $[3.6,5.4]$ for $t$=6.8 s, and $[0,3.4]$
  for $t$=10.3 s. }
\label{fig5}
\end{figure}

Since the focus of this study is closely related to the 
nuclear energy release during strong encounters, practically all 
following calculations refer to cases with penetration factors 
$\beta$ clearly beyond unity, see Table 1. To illustrate the 
evolution of a white dwarf that is just marginally disrupted, we 
show in Fig.~\ref{fig4} a 0.6 \Msun CO white dwarf that passes
a 1000 \Msun black hole with a penetration factor of $\beta=0.9$
(run 10, Table 1; keep in mind that the tidal radius as defined above is approximate
and its exact value depends in the internal structure of the 
disrupted star). The upper panel shows snapshots at t=0.34, 3.43, 
6.86, 10.29, 13.72, 17.15, 20.58 and 24.01 s after the simulation 
start at a separation of 3 $R_\tau$. The lower panel shows late
stages at t= 138.24 and 380.53 s. The circle indicates in both 
cases the location of the tidal radius. At about 1.5 $R_\tau$
the star becomes noticeably deformed and substantially spun up
by the time it reaches pericentre (snapshot 5, upper panel). While 
receeding from the black hole large, puffed-up lobes form at the 
extremes of the star. The inner lobe returns to the black hole 
(see t=138.24 s, lower panel) while the outer is ejected to infinity,
both are connected by a well-defined and homogenous spaghetti-like
tube of white dwarf debris.\\
The opposite limit, an exremely strong encounter, is illustrated in
Figure~\ref{fig5}. The snapshots are taken from our numerical 
simulations of a 0.2 $M_\odot$ white dwarf
approaching a $10^{3}M_\odot$ black hole on a parabolic orbit with
pericenter distance well within the tidal radius ($R_{\rm p}=
R_{\tau}/12$; run 1 in Table 1).  While falling inwards
towards the hole, the white dwarf develops a quadrupole distortion,
which attains an amplitude of order unity by the time of disruption at
$R \sim R_{\tau}$.

When the star is still far from the black hole, it remains close to
its initial stationary equilibrium state as characterized by the usual
virial relation. This is because the timescale characterizing the rate
of change of the tidal force will initially be very long compared with
the intrinsic timescale characterizing the corresponding quadrupole
oscillations of the star. The tidal bulge raised on the star by the
black hole becomes an order unity distortion near the tidal
radius. The resultant gravitational torque spins it up to a good
fraction of its corotation angular velocity by the time it gets
disrupted. The large surface velocities and the order unity tidal
bulge combine to overcome the star's self gravity and lead to the
disruption of the star.

\begin{figure}
\epsscale{0.7}
\plotone{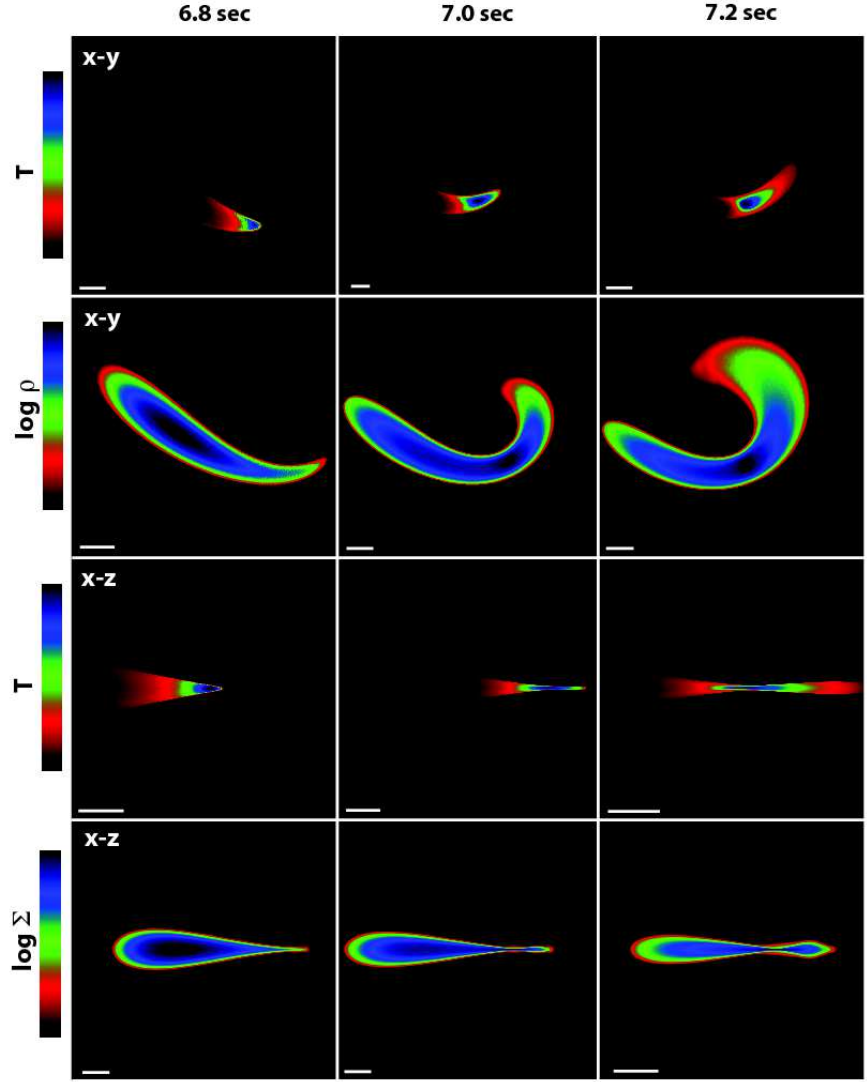}
\caption{Tidal deformation of the white dwarf before and after passage
  through pericenter, as the star attains its maximum degree of
  compression (same simulation as referred to in Figure~\ref{fig5}).
  The panels in the upper rows show cuts of temperature (in units of $10^6$ K)
  and density (in cgs units) through the orbital (xy-)
  plane. Color bar gives the values of $T=[T_{\rm min}, T_{\rm max}]$
  ($\log \rho=[\log \rho_{\rm min}, \log \rho_{\rm max}]$): $[0,330]$
  ($[3.6,5.4]$) for $t$=6.8 s, $[0,1280]$ ($[1.0,6.0]$) for $t$=7.0 s,
  and $[0,3200]$ ($[1.0,6.0]$) for $t$=7.2 s. The panels in the lower two
  rows show the temperature (in units of $10^6$ K) and column density
  (in cgs units) distributions in the xz-pane (averaged along the
  y-direction). Color bar gives the values of $T=[T_{\rm min}, T_{\rm
      max}]$ ($\log \Sigma=[\log \Sigma_{\rm min}, \log \Sigma_{\rm
      max}]$): $[0,220]$ ($[12.0,15.7]$) for $t$=6.8 s, $[0,720]$
  ($[12.0,16.0]$) for $t$=7.0 s, and $[0,1700]$ ($[12.0,16.5]$) for
  $t$=7.2 s. The dimension of the bar scale is $10^{9}$ cm. }
\label{fig6}
\end{figure}

Our present investigation is concerned with cases of deeper
penetration within the tidal radius so that the core as well as the
envelope are affected. The behavior of a white dwarf passing well
within the tidal radius exhibits special features
\citep{cl1983,b1983}.  As illustrated in Figure~\ref{fig6}, the
degenerate star is not only elongated along the orbital direction but
also severely compressed perpendicular to the orbital plane. This
anisotropy can be understood as arising from the fact that the
principal tidal axes within the orbital plane will rotate prior to
pericenter passage so that the corresponding effects of elongation and
compression will roughly cancel out, whereas the third principal axis
retains a fixed direction so that compression orthogonal to the
orbital plane is uncontested (see, e.g., the appendix of
\citep[][]{bl2008}). Each section of the star is squeezed
through a point of maximum compression at a fixed point on the star's
orbit. This takes place on a timescale comparable to the crossing time
of the star through periastron, 
\be
 \delta t \sim {R_{\rm wd}
  \over v_{\rm p}} \simeq 0.2 \left({M_{\rm wd} \over 0.6
  M_\sun}\right)^{-1/6} \left({R_{\rm wd} \over 10^{9}{\rm
    cm}}\right)^{3/2} \left({M_{\rm BH} \over 10^3\;M_\sun
}\right)^{-1/3}\;{\rm s}, \ee where \be v_{\rm p}\sim (R_{\rm
  g}/R_{\tau})^{1/2} c \simeq 5 \times 10^{9} \left({M_{\rm wd} \over 0.6
  M_\sun}\right)^{1/6} \left({R_{\rm wd} \over 10^{9}{\rm
    cm}}\right)^{-1/2} \left({M_{\rm BH} \over 10^3\;M_\sun
}\right)^{1/3}\; {\rm cm\;s^{-1}},
\ee
is the orbital velocity at periastron.

\begin{figure}
\epsscale{1.0}
\plotone{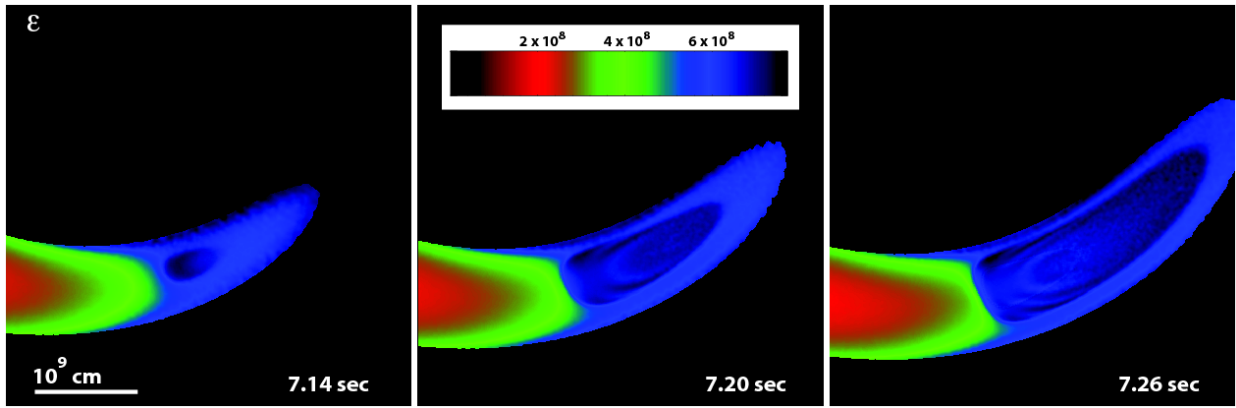}
\caption{Evolution of the entropy $\varepsilon$ of the central portion
  of the disrupted white dwarf just after it attains its maximum
  degree of compression (same simulation as referred to in
  Figures~\ref{fig5} and ~\ref{fig6}). This compression is halted by a
  shock, raising the matter to a higher adiabat.  The panels show
  entropy (in cgs units) cuts through the orbital (xy-) plane.}
\label{fig7}
\end{figure}

During this very short lived phase, the star attains its maximum
degree of compression (here by a factor $\sim 100$) although its
orbital-plane section increases only by a factor of a few due to the
continually changing directions of the two principal tidal axis
(Figure~\ref{fig6}). The distortions $\Delta R_{\rm wd}/R_{\rm wd}
\sim 1$ impart supersonic bulk flow velocities of order \be {R_{\rm
    wd} \over \delta t} \sim v_{\rm p} \ll c_s \sim 2.8 \times
10^{8}\left({R_{\rm wd} \over 10^{9}{\rm cm}}\right)^{-1/2}
\left({M_{\rm wd} \over 0.6\;M_\sun }\right)^{1/2}\; {\rm
  cm\;s^{-1}}\ee during the drastic compression of the stellar
material. This compression is thus halted by a shock
\citep{ko2004,bl2008}, raising the matter, which then rebounds
perpendicular to the orbital plane, to a higher adiabat
(Figure~\ref{fig7}). As a result, the temperatures increase sharply
and trigger explosive burning (of He for the case shown in
Figure~\ref{fig6}).

The typical temperature of the shocked star and the thermal energy
produced by shock heating can be roughly estimated from 
the virial theorem as 
\be
T \sim 6 \times 10^{8} \left({R_{\rm wd} \over 10^{9}{\rm cm}}\right)^{-1}
\left({M_{\rm wd} \over 0.6\;M_\sun }\right)\; {\rm K},\ee and \be
E_{\rm therm} \sim E_{\rm G} \sim 10^{50} \left({R_{\rm wd} \over 10^{9}{\rm
    cm}}\right)^{-1} \left({M_{\rm wd} \over 0.6\;M_\sun }\right)^2\;
{\rm erg}.
\ee 
Adiabatic compression alone can only increase the
stellar surface and interior layer temperatures by a modest factor.
If adiabatic compression was the only source of heating, the response
to the flow to the varying potential would be $\sim R_{\rm wd}/c_s\sim
4$ s, which is significantly longer than the dynamic crossing time.

\begin{figure}
\epsscale{1.0}
\plotone{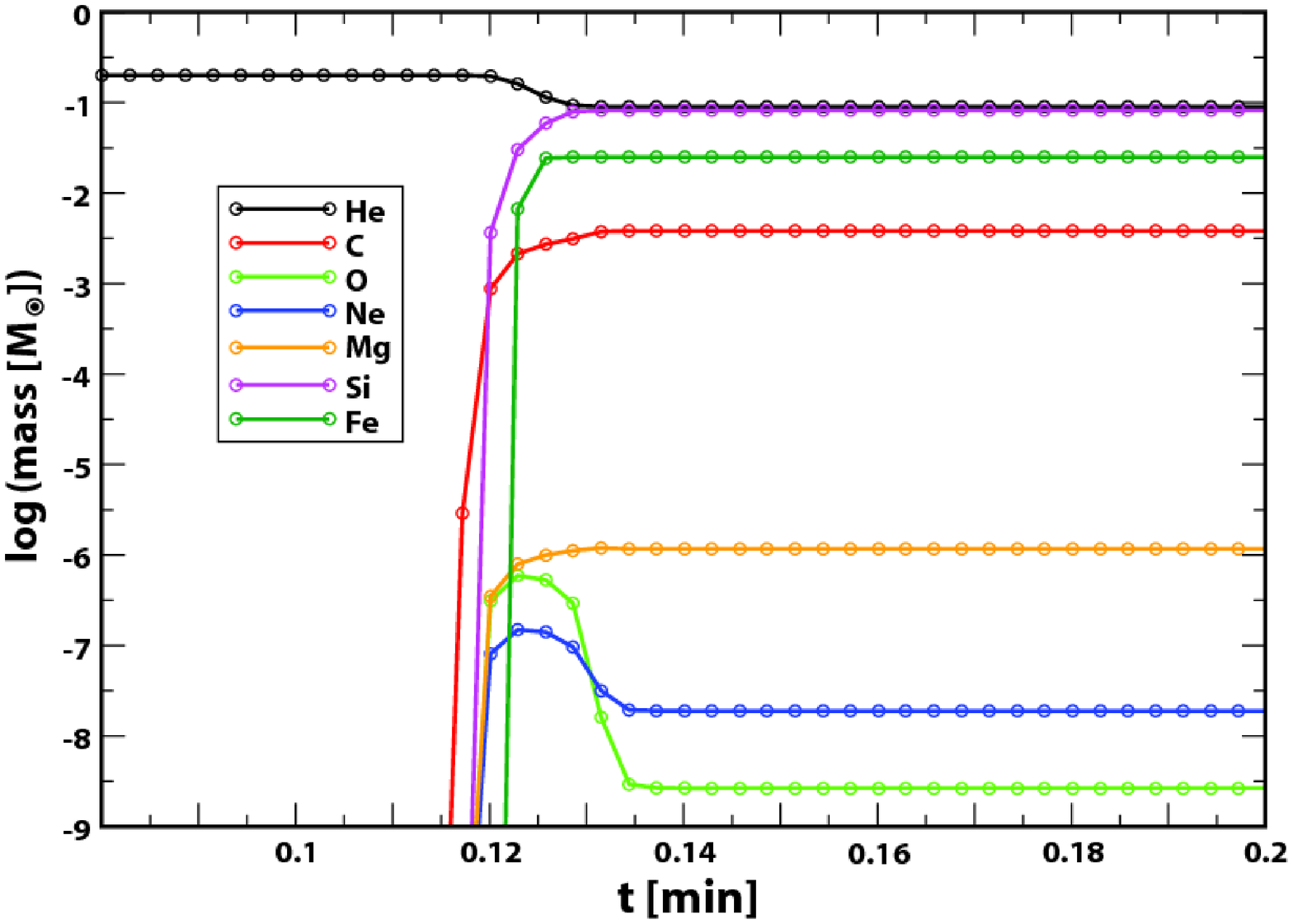}
\caption{Evolution of the abundances during the disruption and
  ignition of a 0.2 $M_\odot$ white dwarf passing a $10^3 M_\odot$
  black hole (same simulation as referred to in Figures
  \ref{fig5}--\ref{fig7}).}
\label{fig8}
\end{figure}

The evolution of the abundances (for the simulation shown in Figure
\ref{fig5}) is shown in Figure~\ref{fig8}, within the 12 s during
which the entropy increases the fastest. During the about one second
long period of compression, the temperature increases beyond $3 \times
10^9$K, approaching but not quite reaching nuclear statistical
equilibrium (NSE). During this brief period of compression, nuclei up
to and beyond Si are synthesized.  The initial composition was pure He
and the final mass fraction in iron-group nuclei is about 15\%. This
result should be taken as a modest underestimate, since the 
seven species nuclear network only provides an approximation 
to the detailed nuclear processes.  Post-processing
calculations, using a 300 isotope nuclear network over thermodynamic
particle histories resulting from these calculations, show 
significant nuclear flow above silicon, for helium-rich portions of 
the gas with peak temperatures above $2 \times10^9$K. As a result, 
heavier elements (like calcium, titanium and chromium) would be made, 
accompanied by a modest increase in the energy generation.  It is thus 
safe to conclude that the white dwarf is tidally ignited and that a 
sizable mass of iron-group nuclei is injected into the outflow.

\begin{figure}
\plotone{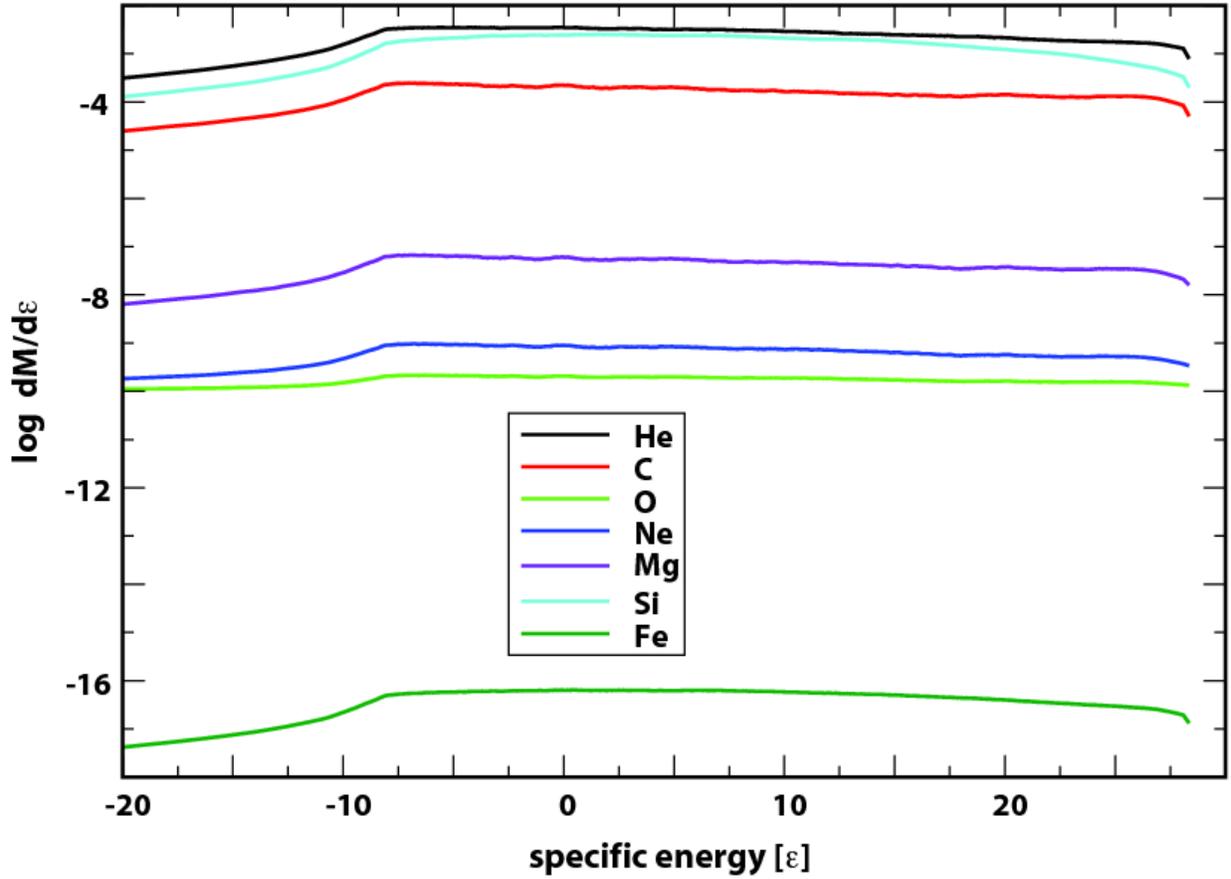}
\caption{Differential mass distributions in specific energy for the
  0.2 $M_\odot$ white dwarf debris (same simulation as referred to in
  Figures \ref{fig5}--\ref{fig8}).}
\label{fig9}
\end{figure}

The variation of the specific energy in the released gas, in the
absence of explosive energy input, is determined mainly by the
relative depth of a mass element across the disrupted star in the
potential well of the black hole \citep{re1988}. This is much 
larger than the binding
energy and the kinetic energy generated by spin-up near pericenter.
Even though the mean specific binding energy of the debris to the hole
is comparable with the self-binding energy of the original star, the
spread about this mean is larger by a factor $(M_{\rm BH}/M_{\rm
  wd})^{1/3}$. Nuclear energy released during the drastic compression
and distortion of the stellar material, further enhances the spread in
specific energies at pericenter (Figure \ref{fig9}), and, as a result,
the mass fraction escaping on hyperbolic orbits is increased from
$\sim$ 50\% to $\sim$ 65\% of the initial mass of the white dwarf. The
mass fraction that is ejected rather than swallowed, though less
spectacular than typical Type Ia supernovae \citep{hi2000}, should
have many distinctive observational signatures (the reader is referred
to Kasen et al. 2008 for a detailed description of the optical light
curves and spectra resulting from the unbound debris before it becomes
translucent).  First, the explosion itself should be different, since
the disrupted, degenerate stars should be, on average, lighter than
those exploding as type Ia supernovae. Second, the spectra should
exhibit large Doppler shifts, as the ejected debris would be expelled
with speeds $\ge 10^4$ km/s. Finally , the optical light curve should
be rather unique as a result of the radiating material being highly
squeezed into the orbital plane (one thus expects different timescales
for conversion of nuclear energy to observable luminosity when
compared with normal type Ia events).

Although the explosion will increase the fraction of ejected debris,
a good fraction ($\sim$ 35\%) remains to be accreted on to 
the hole.  The returning gas does
not immediately produce a flare of activity from the black hole. First
material must enter quasi circular orbits and form an accretion
torus \citep{re1988,ev1989}. Only then will viscous effects release
enough binding energy to power a flare. The bound orbits are very
eccentric, and the range of orbital periods is large. For white
dwarfs, the orbital semi-major axis of the most tightly bound debris
is \be a \sim 300 \left({M_{\rm BH} \over 10^3\;M_\sun }\right)^{-1/3}
\left({R_{\rm wd} \over 10^{9}{\rm cm}}\right) \left({M_{\rm wd} \over
  0.6\;M_\sun }\right)^{-2/3}\;R_{\rm g}, \ee and the period is only
\be t_{\rm a} \sim 150 \left({a \over 300 R_{\rm g}}\right)^{3/2}
\left({M_{\rm BH} \over 10^3\;M_\sun }\right)^{-1/2}\;{\rm s}.\ee

\begin{figure}
\plotone{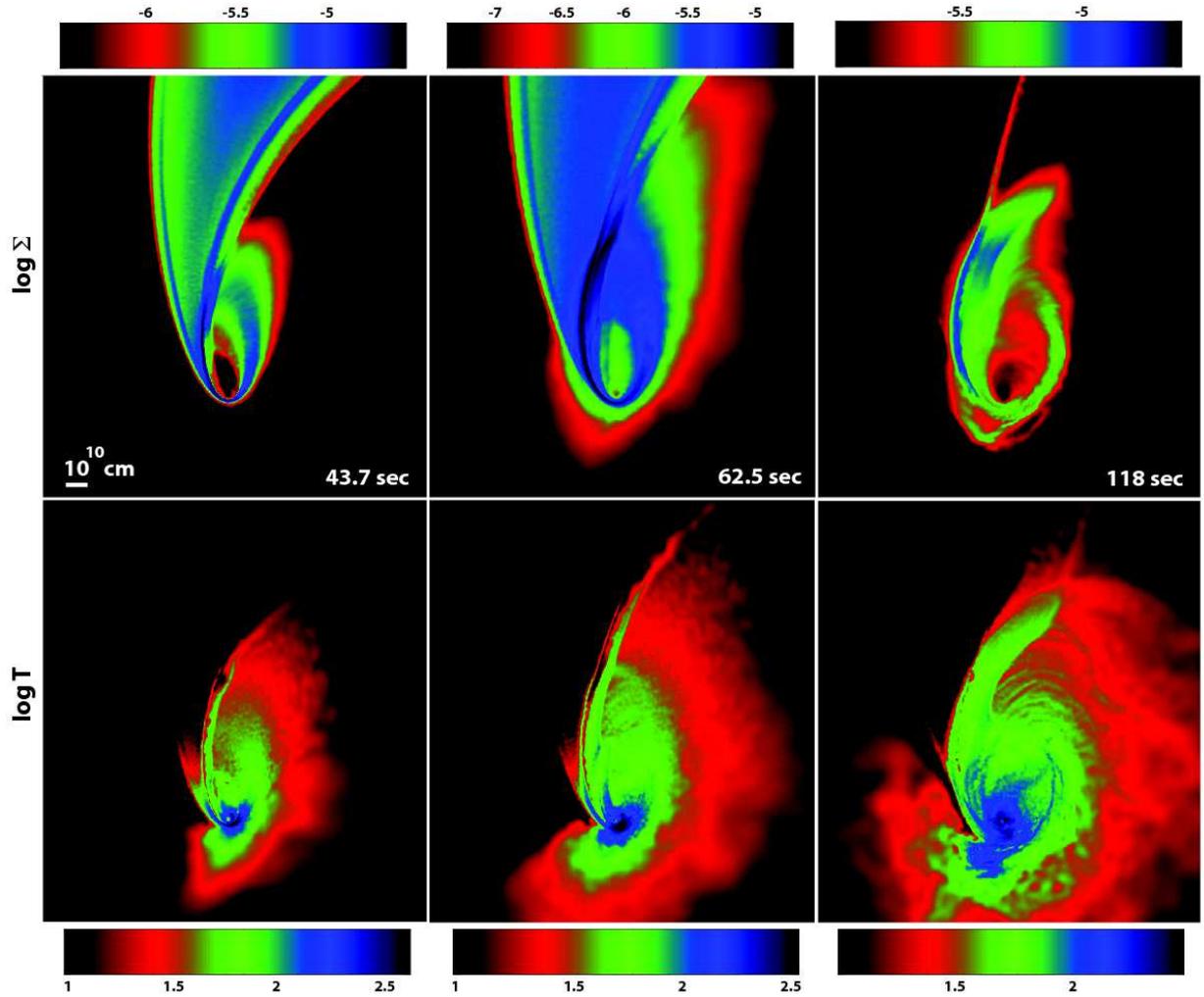}
\caption{Column density (in cgs units) and temperature (in units of
  $10^6$ K) distributions in the orbital plane of the bound debris
  minutes after disruption. The most tightly bound debris would
  transverse an elliptical orbit with major axis $\sim 300 R_{\rm g}$
  before returning to $R\approx R_{\tau}$ (same simulation as referred
  to in Figures \ref{fig5}--\ref{fig9}). }
\label{fig10}
\end{figure}

\begin{figure}
\epsscale{1.0}
\plotone{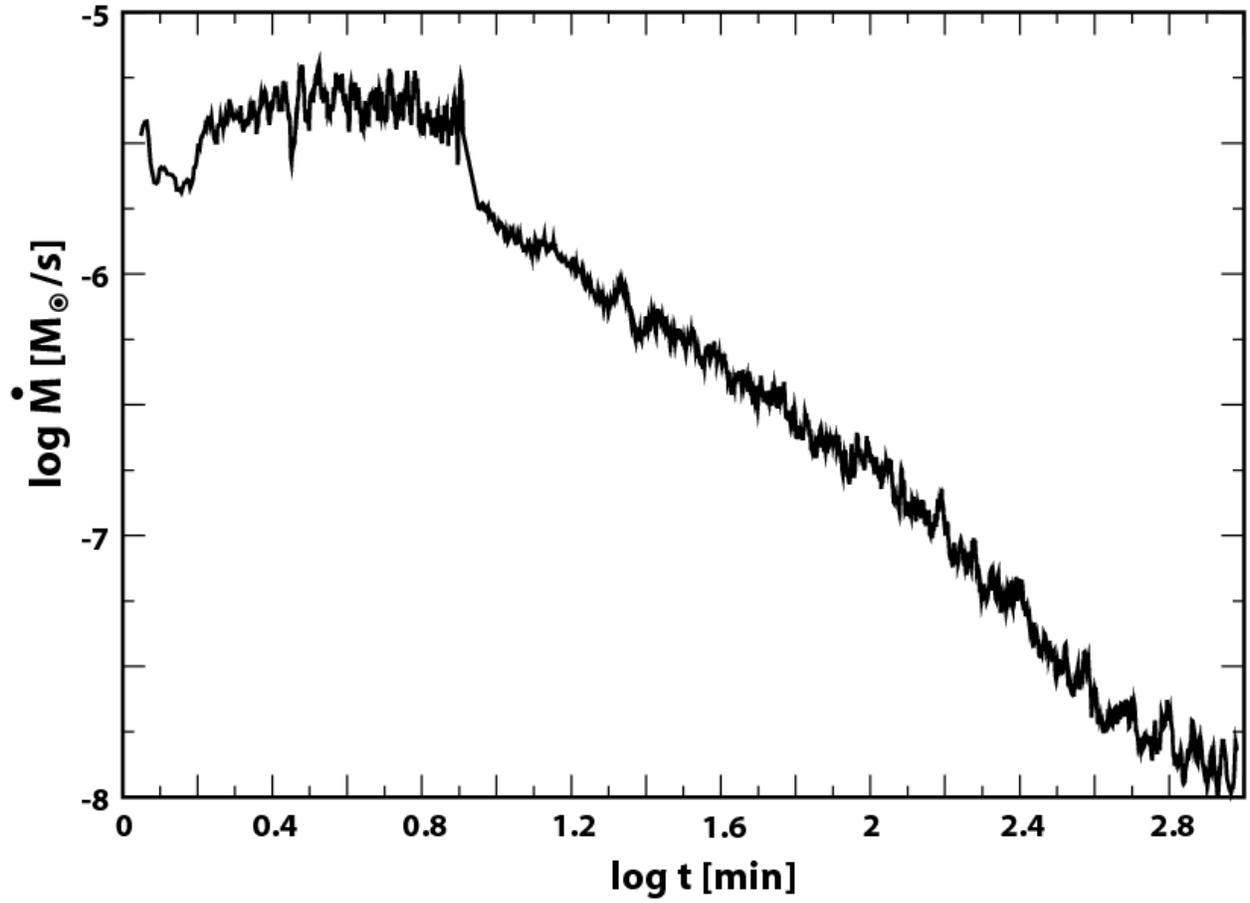}
\caption{The rate at which the 0.2 $M_\odot$ white dwarf debris
  returns to the vicinity of the black hole (same simulation as
  referred to in Figures \ref{fig5}--\ref{fig10}).}
\label{fig11}
\end{figure}

If the gaseous debris suffered no internal dissipation due to high
viscosity or shocks, it would, after one or two orbital periods, form
a highly elliptical disc with a big spread in apocenter distances
between the most and least bound orbits, but where at pericenter,
$R_{\rm p}$, the radial focusing of the orbits acts as an effective
nozzle (Figure~\ref{fig4} and \ref{fig10}). After pericenter passage, the outflowing
gas is on orbits which collide with the infalling stream near the
original orbital plane at apocenter, giving rise to an angular
momentum redistributing shock (Figure~\ref{fig10}) much like those in
cataclysmic variable systems. The debris raining down would, after
little more than its free-fall time, settle into a disc. This orbiting
debris starts to form when the most tightly bound debris falls
back. The simulation shows that the first material returns at a time
$\leq t_{\rm a}$, with an infall rate of about $10^2 M_\odot\;{\rm
  yr^{-1}}$ (Figure~\ref{fig11}). Such high infall rates are expected
to persist, relative steadily, for at least a few orbital periods,
before all the highly bound material rains down. The vicinity of the
hole would thereafter be fed solely by injection of the infalling
matter. The early mass infall rate is sensitive to the
stellar structure \citep{lodato08,ramirezruiz08}, at late times,
$t \geq t_{\rm fb} \approx 600$ s, it drops off as $t^{-5/3}$
\citep{re1988,phinney89}. Once the torus is formed, it will evolve
under the influence of viscosity, radiatively cooling winds and time
dependent mass inflow.

A luminosity comparable to the Eddington value, 
$\sim L_{\rm Edd}=10^{41}(M_{\rm BH}/10^3M_\odot)\;{\rm erg\;s^{-1}}$, 
can therefore only be maintained for at most a year;
thereafter the flare would rapidly fade. It is clear that most of the
debris would be fed to the hole far more rapidly than it could be
accepted if the radiative efficiency were high; much of the bound
debris must either escape in a radiatively-driven outflow or be
swallowed inefficiently. The rise and the peak bolometric luminosity
can be predicted with some confidence. However, the effective surface
temperature (and thus the fraction of luminosity that emerges
predominantly in the soft X-ray band) is harder to predict, as it
depends on the size of the effective photosphere that shrouds the
hole. Such transient signals, if detected, would be a compelling testimony
for the presence of moderately massive black holes in the centers of
globular clusters and dwarf galaxies.

\section{Influence of Black Hole's Mass on Disruption}\label{bh}
\begin{figure}
\epsscale{1.0}
\plotone{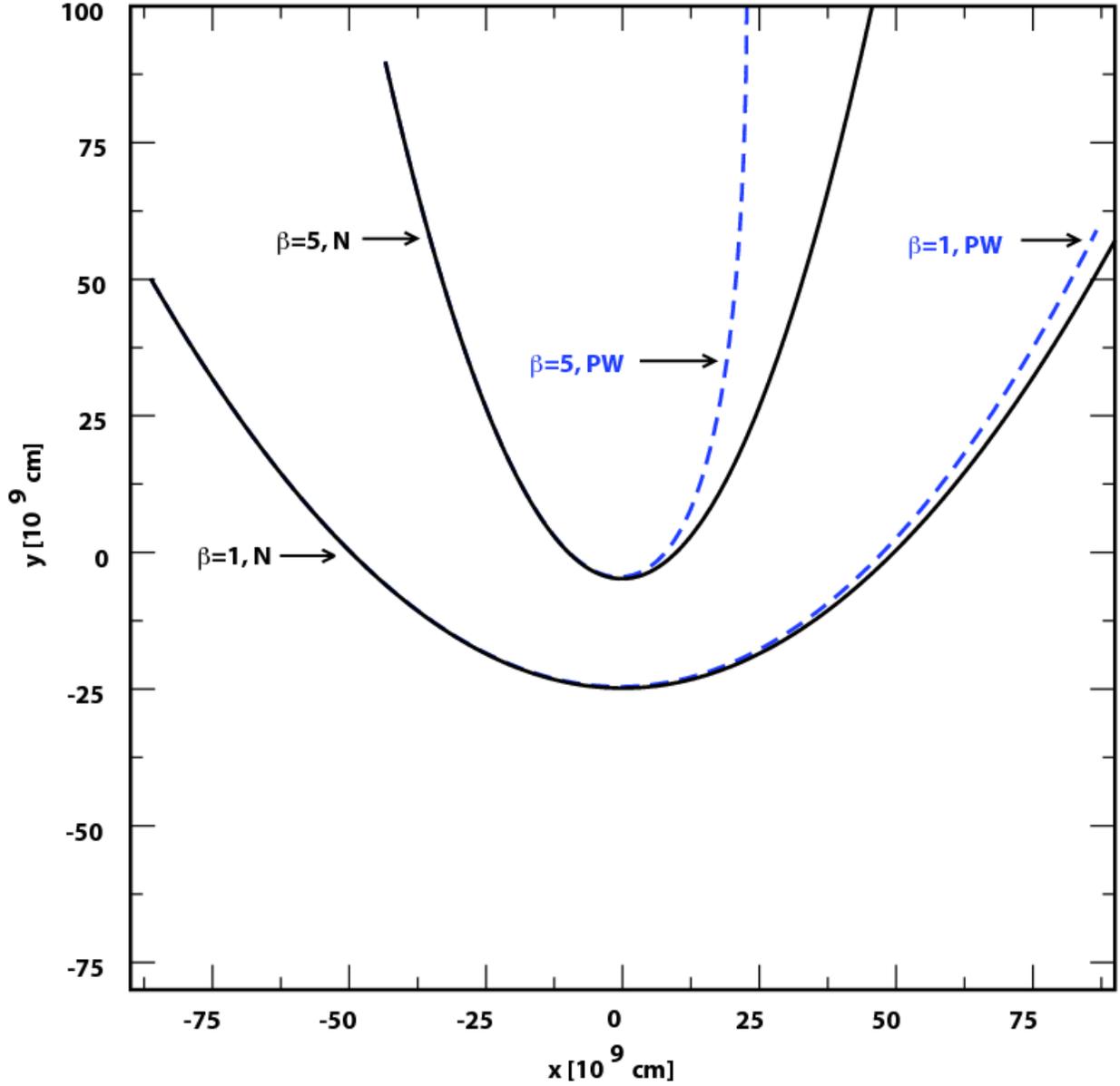}
\caption{The importance of relativistic effects from the central
  black hole in determining the behavior of white dwarfs passing
  within the tidal radius.  The trajectory of the center of mass of a
  0.2 $M_\odot$ white dwarf approaching a $10^3\;M_\odot$ black hole
  on a parabolic orbit with impact parameter $\beta$ are shown using 
  Newtonian gravity (N; black) and the pseudo-Newtonian relativistic
  potential of Paczy\'nski-Wiita (PW, blue). Each time, the star is
  modeled with more than $10^5$ partices.}
\label{fig12}
\end{figure}

The characteristic tidal radius, $R_\tau$, for a given white dwarf star
is solely determined by the black hole's mass, while the
strength of the tidal encounter is traditionally measured by the
dimensionless parameter $\beta=R_\tau/R_{\rm p}$ (provided $\beta \leq
\beta_{\rm g}$). The aim of this section is to illustrate, with the help
of a few specific calculations, the role of the black hole's mass in
shaping the evolution and ignition of the disrupted white dwarf. To
quantify this, we have performed detailed three-dimensional,
hydrodynamical calculations of the dynamics of a $0.2 M_\odot$ white
dwarf approaching black holes of various masses on parabolic orbits
with $\beta=5$.  For black hole masses $10^2 M_\odot \le M_{\rm BH} \le
10^4 M_\odot$ and $\beta=5$, a $0.2 M_\odot$ white dwarf can be
disrupted without entering the strong relativistic domain
(Figure~\ref{fig1}). However, complete disruption occurs sufficiently
close to the hole for a Newtonian approximation to be inadequate
(Figure~\ref{fig12}).

\begin{figure}
\epsscale{0.7}
\plotone{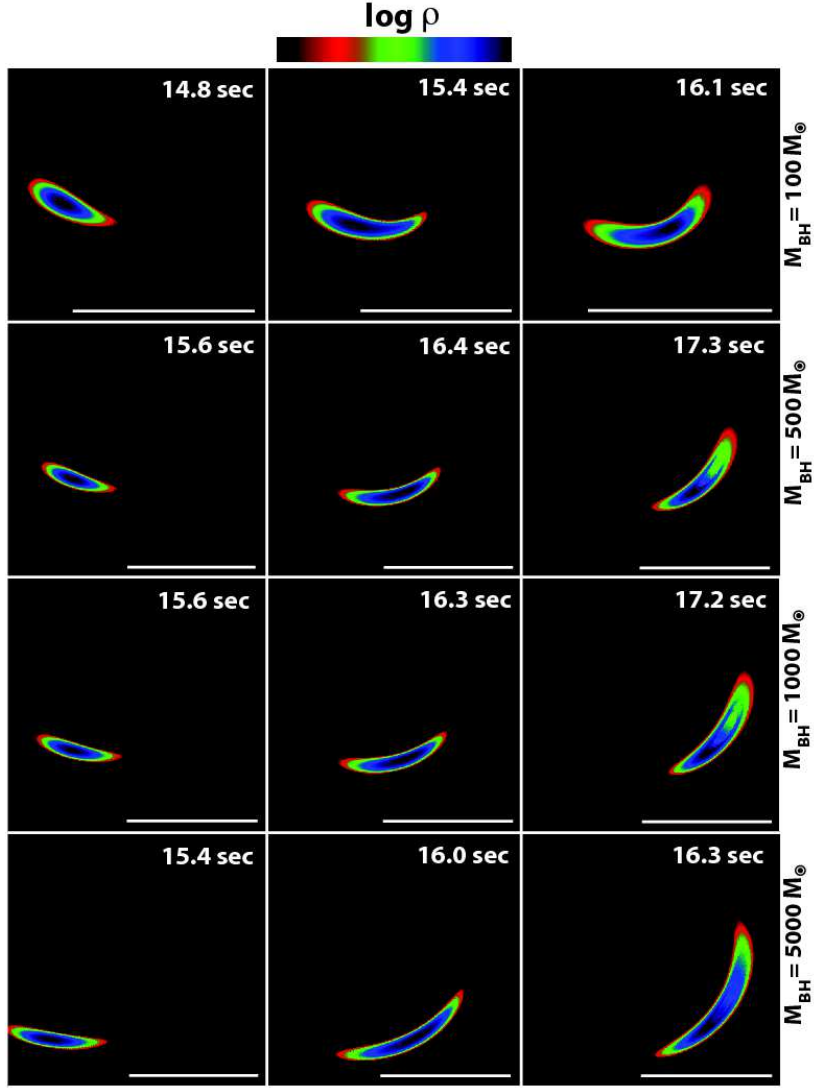}
\caption{A 0.2 $M_\odot$ white dwarf approaching black holes of
  various masses on a parabolic orbit with $\beta=5$. The panels in
  the show density (in cgs units) cuts through the orbital (xy-)
  plane. Color bar gives the values of $\log \rho=[\log \rho_{\rm
      min}, \log \rho_{\rm max}]$.  For $M_{\rm BH}=10^2M_\odot$, $\log
  \rho=$ $[4.25,5.3]$ for t$=14.8$ s, $[4.0,5.4]$ for t$=15.4$, and
  $[3.75,5.8]$ for t$=16.1$. For $M_{\rm BH}=5 \times 10^2M_\odot$,
  $\log \rho=$ $[4.15,5.3]$ for t$=15.6$ s, $[3.9,5.65]$ for t$=16.4$,
  and $[3.75,5.8]$ for t$=17.3$.  For $M_{\rm BH}= 10^3M_\odot$, $\log
  \rho=$ $[4.1,5.4]$ for t$=15.6$ s, $[3.9,5.65]$ for t$=16.3$, and
  $[3.4,5.6]$ for t$=17.2$.  For $M_{\rm BH}= 5 \times 10^3M_\odot$,
  $\log \rho=$ $[4.0,5.4]$ for t$=15.4$ s, $[3.7,5.75]$ for t$=16.0$,
  and $[3.4,5.8]$ for t$=16.3$.}
\label{fig13}
\end{figure}

The behavior of a $0.2 M_\odot$ white dwarf approaching black holes of
various masses is illustrated in Figure~\ref{fig13}.  As discussed
previously, the effects of the black hole's mass (for a fixed
$\beta$ and $M_{\rm wd}$) will be to move the critical pericentric
distance by $M_{\rm BH}^{1/3}$ and change the star's crossing time through the
maximum compression point by $M_{\rm BH}^{-1/3}$. As a result, stars
approaching black holes of increasing mass will appear more elongated
along the orbital plane, as each section of the star is squeezed
faster through a point of maximum compression at a fixed point on the
star's increasingly extended orbit. Such stars are not only more
elongated but are even more severely compressed into a prolate
shape. If the initial white dwarf is made of pure He (as in the 0.2
$M_\odot$ case), the combustion rate will be determined by the
$3\alpha$ reaction, on time scale approximately given by \citep{kh1986}
\be 
t_{\rm b}
\approx 9.0 \times 10^{-4} T_9^3 \exp(4.4/T_9) \; \rho_6^{-2}, 
\label{eq:t_burn}
\ee
where $T_9$ is the temperature in units $10^9$ K and $\rho_6$ the
density in $10^{6}$ \gcc.

\begin{figure}
\epsscale{1.0}
\plotone{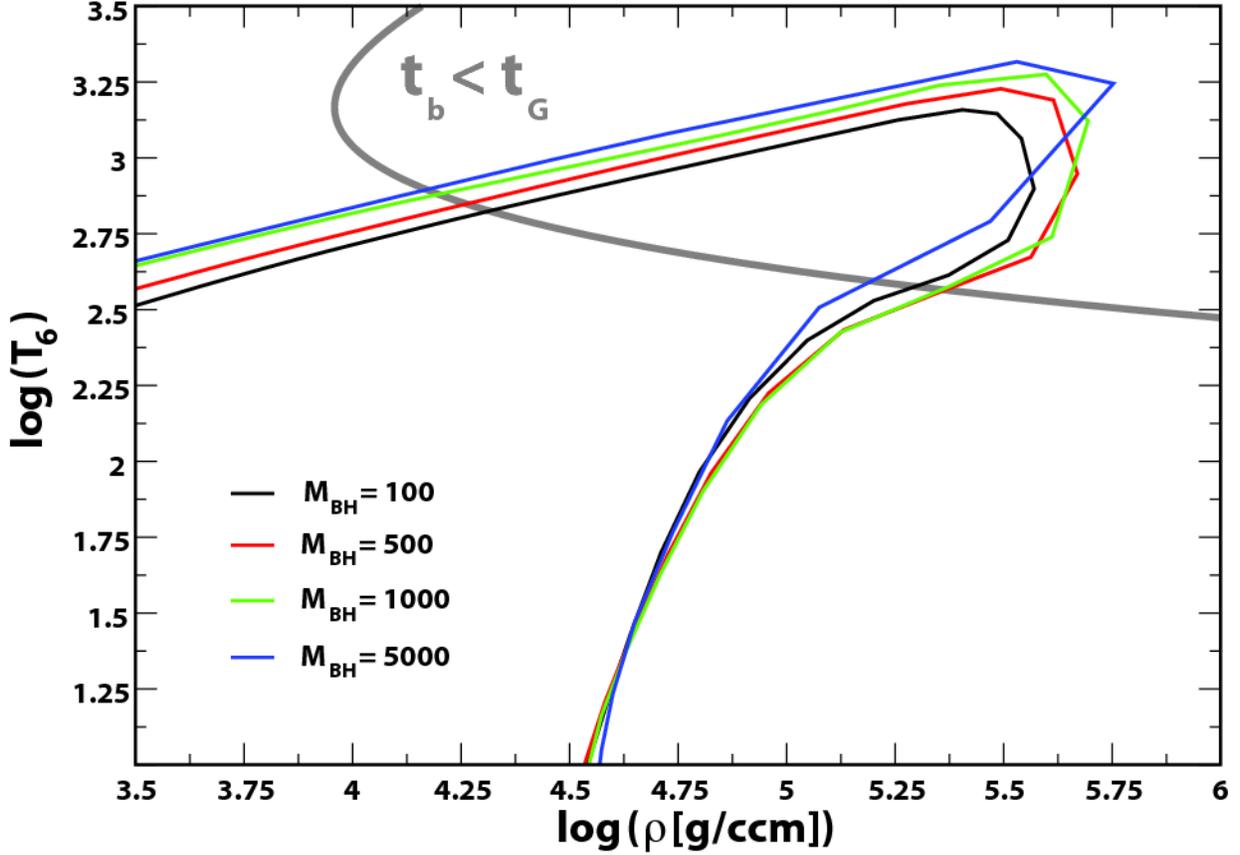}
\caption{The importance of the central black hole's mass in
  determining the behavior of a 0.2 $M_\odot$ white dwarf passing
  within $\beta=5$ (same simulations as referred to in
  Figure~\ref{fig13}). The evolution of the compressed, and tidally
  disrupted white dwarf in the $\rho-T$ plane. The hottest 10\% of the
  particles are identified and their average temperature (in units of
  $10^6$ K) is plotted as a function of their average density (in cgs
  units).  These trajectories always start cold and dense (right lower
  corner) and become hot and during the black hole flyby. If the time
  scale on which the white dwarf can react (dynamical time scale 
  $t_{\rm G}$) is longer than the burning time scale ($t_{\rm b}$, see 
  Eq.~(\ref{eq:t_burn})), the star cannot expand rapidly enough to quench 
  burning.}
\label{fig14}
\end{figure}

\begin{figure}
\epsscale{1.0}
\plotone{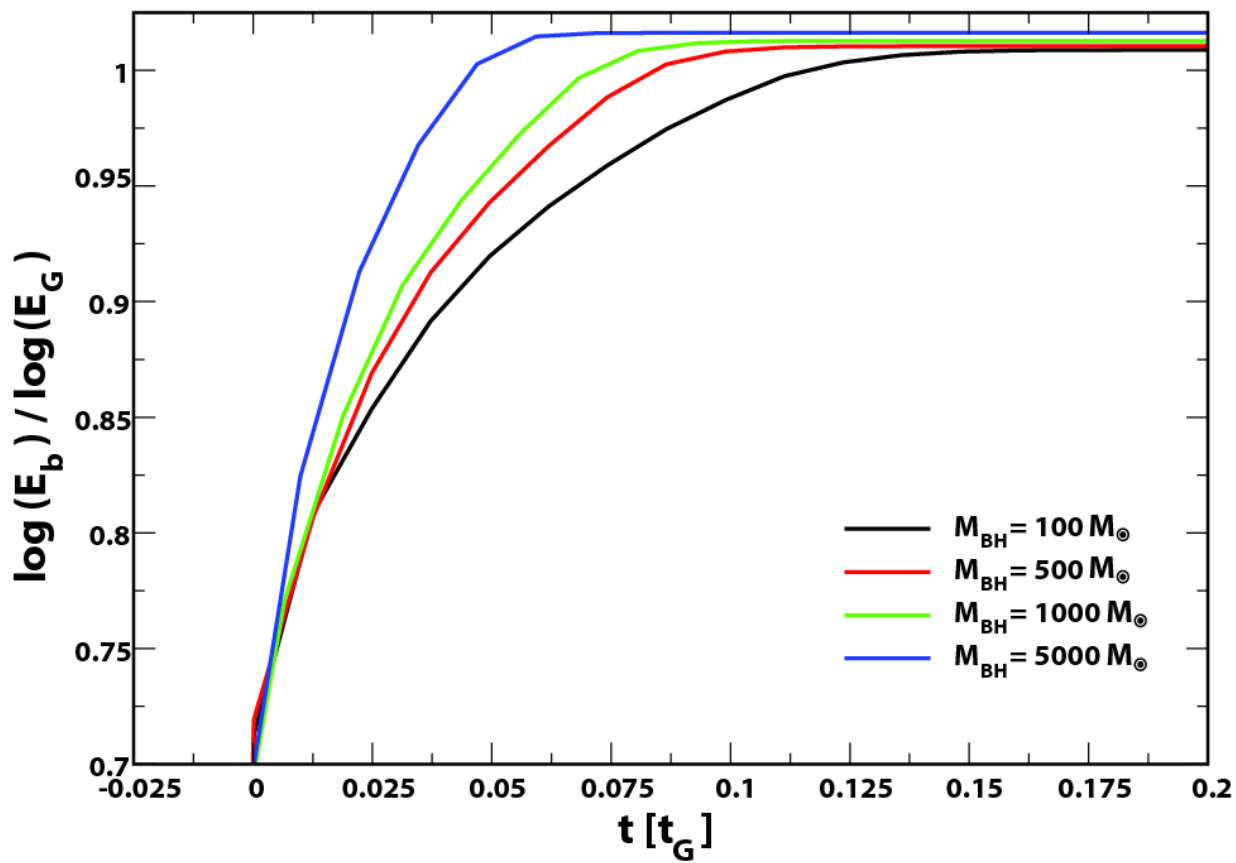}
\caption{Energy generated in nuclear burning (in units of the star's
  binding energy) as a function of time (same simulations as referred
  to in Figures~\ref{fig13} and \ref{fig14}). Here time is measured in
  units of the dynamical timescale ($t_{\rm G}$) of the initial 0.2
  $M_\odot$ white dwarf. In addition, the time axis has been shifted
  so that the maximum $E_{\rm b}$ occurs at $t=0$.}
\label{fig15}
\end{figure}

If the time scale on which the white dwarf can react, its dynamical
time scale $t_{\rm G}= (G\bar{\rho}_{\rm wd})^{-1/2}$, is much shorter
than the burning time scale $t_{\rm b}$, the star can expand rapidly
enough to quench burning by a reduction of temperature and
density. Appreciable burning will therefore only take place if $t_{\rm
  b} \ll t_{\rm G}$. As illustrated in Figure~\ref{fig14}, this
comparison of time scales can be used as a simple estimate for whether
or not substantial He-burning will occur. Promising models for
thermonuclear explosions are those that reach high temperatures at
high densities (upper right corner). For a fixed $\beta$, 
increasing $M_{\rm  BH}$ raises the nuclear energy release.  
For black hole masses $M_{\rm BH}>10^2$ \msun, nuclear energy
released during the drastic compression and distortion of a 0.2
$M_\odot$ white dwarf approaching moderately massive black holes with
$\beta=5$ is fast enough to release energy in excess of that required
to tear the star apart: $E_{\rm G}$ (Figure~\ref{fig15}). When the
black hole mass is $\gg 10^4 M_\odot$ most white dwarfs would be,
however, swallowed whole ($R_ \tau<R_{\rm g}$).

\section{Tidal Disruption for a Variety of White Dwarfs}\label{var}
White dwarfs are thought to be the final evolutionary state of all
stars whose mass is below $\sim 10$ \Msun \citep{w1990}. 
Stars with masses less than about half a solar mass become
degenerate before helium ignition, and therefore will end 
their lifes as helium white dwarfs. In isolation, such low 
mass stars have life times much longer than the present age 
of the Universe. Still, such He white dwarfs are observed, 
they are thought to be the result of the evolution of a close 
binary system \citep[e.g.][]{he2002}.
Stars of low or medium mass, achieve helium burning and become 
carbon-oxygen white dwarfs (perhaps the commonest sort).  
The observed white dwarf mass distribution is
strongly peaked around 0.6 \Msun \citep{ke2007}.\\
When a white dwarf is subject to strong tidal compression,
triggering of nuclear processes in the stellar cores depends on
initial composition.  To investigate this dependence, we have 
performed calculations with various white dwarf masses/compositions.
As outlined above, our white dwarf models are initialized with pure 
He-composition for $M_{\rm wd} < 0.6$ \Msun and 50\% carbon and 50\% oxygen 
otherwise.

\begin{figure}
\epsscale{1.0}
\plotone{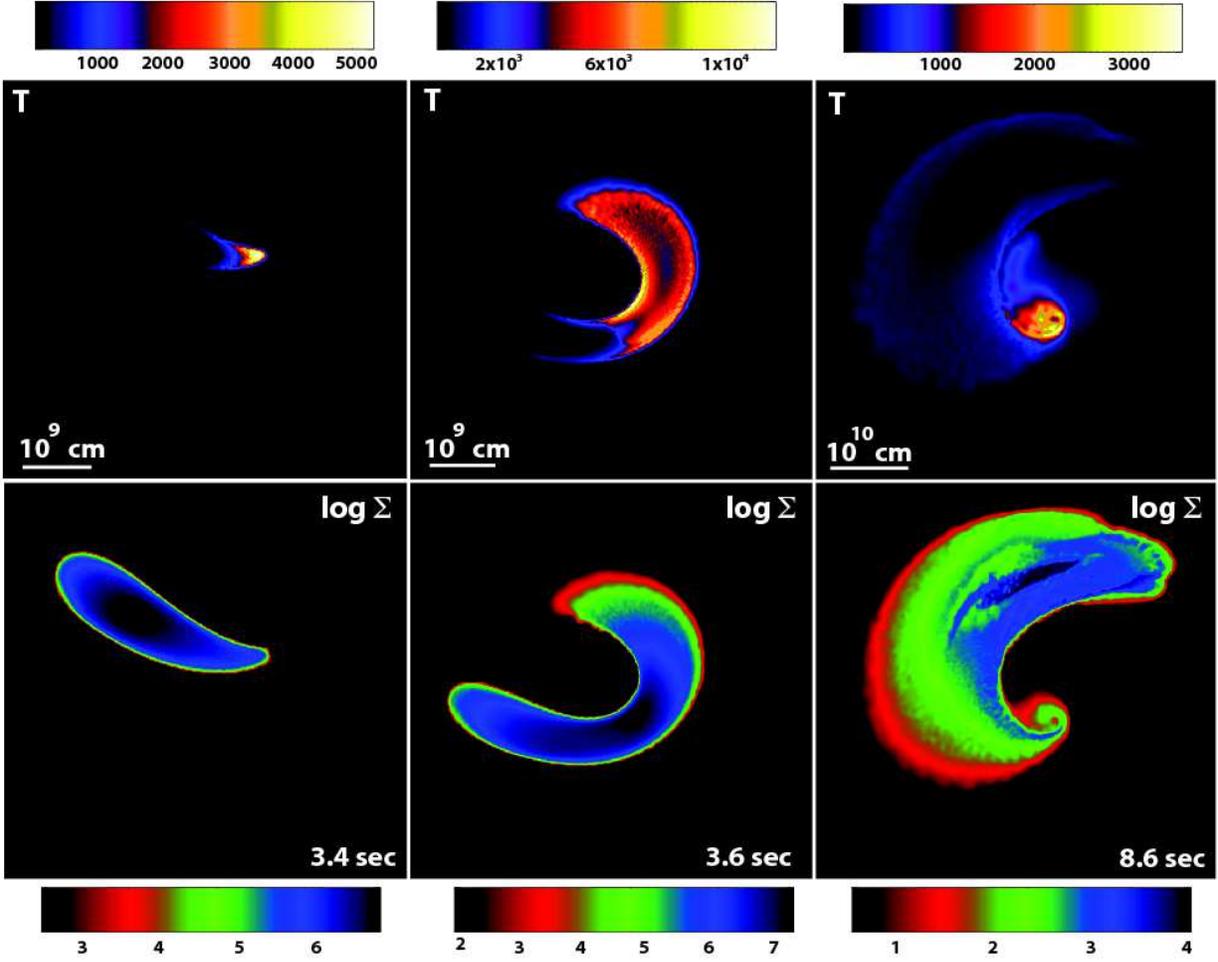}
\caption{A 1.2 $M_\odot$ white dwarf (modeled with more than $5 \times
  10^5$ SPH particles) approaching a $500\;M_\odot$ black hole on a
  parabolic orbit with pericenter distance $r_{\rm min} = r_{\rm
    T}/3.2$. Shown are temperature (in units of $10^6$ K) and surface
  density (in cgs units) cuts at various instants along the orbital
  (xy-) plane. }
\label{fig16}
\end{figure}

\begin{figure}
\epsscale{1.0} 
\plotone{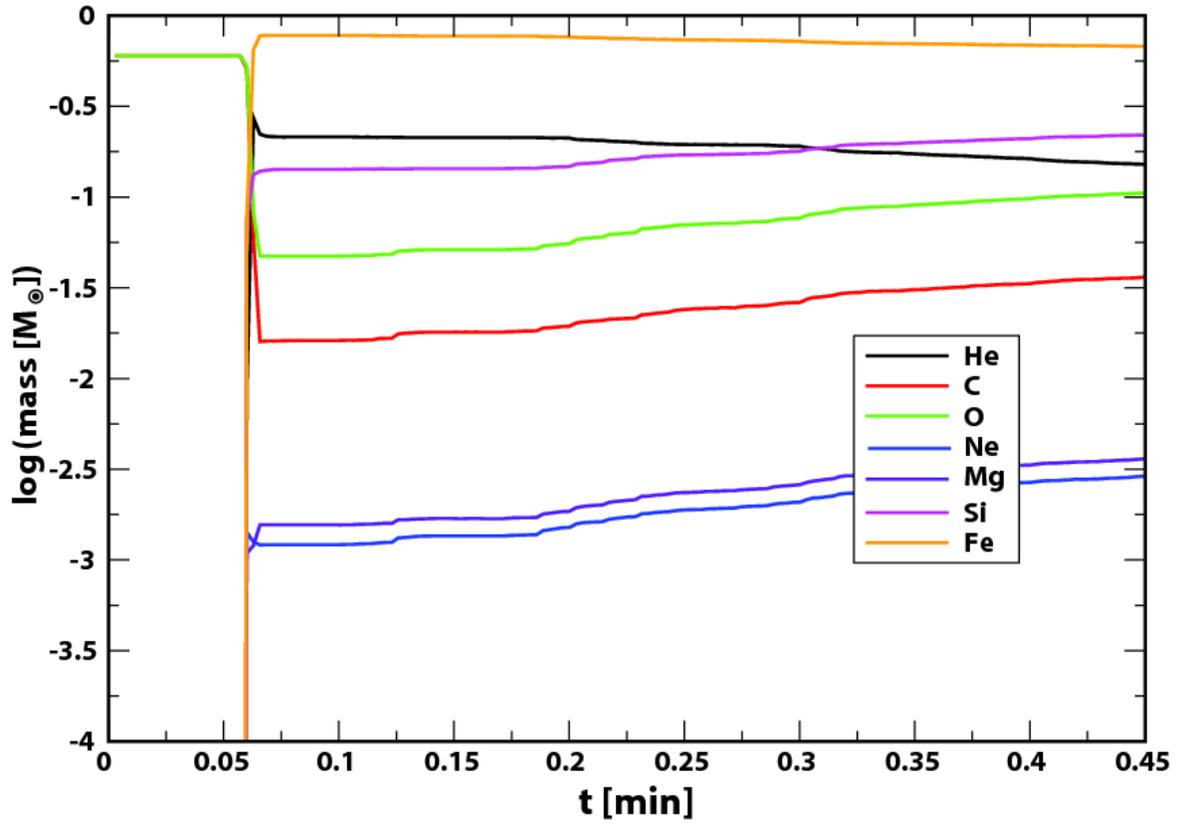}
\caption{Evolution of the abundances during the disruption and
  ignition of a 1.2 $M_\odot$ white dwarf approaching a $500 M_\odot$
  black hole (same simulation as referred to in Figure \ref{fig16}).}
\label{fig17}
\end{figure}

Although the thermodynamical evolution and the nuclear energy release
are very sensitive to the initial stellar composition (in this text,
we have so far considered only a pure $^4$He 0.2 $M_\odot$ 
white dwarf), we
found that ignition is in fact a natural outcome for white dwarfs of
all masses passing well with in the
tidal radius. For example, a C/O $1.2 M_\odot$ white dwarf (here
assumed to be composed of 50\% carbon and 50\% oxygen throughout the
star) approaching a 500 $M_\odot$ black hole on a parabolic orbit with
pericenter distance $\beta = 3.2$ ignites (Figure~\ref{fig16}) and, as
a result, at least 0.6 $M_\odot$ of Fe are synthesised in the flow
(Figure~\ref{fig17}). Similarly, at least 0.66 $M_\odot$ of Fe are
synthesised when a C/O $1.2 M_\odot$ white dwarf approaches a $10^3
M_\odot$ black hole with $\beta = 3$.  By contrast with the previous
two cases, explosive energy release is too slow to release much energy
on a dynamical timescale when a C/O $1.2 M_\odot$ white dwarf
approaches a $10^3M_\odot$ black hole with $\beta =1.5$ and, as a
result, only $1.4 \times 10^{-2} M_\odot$ of Fe are synthesised.

\begin{figure}
\epsscale{1.0} 
\plotone{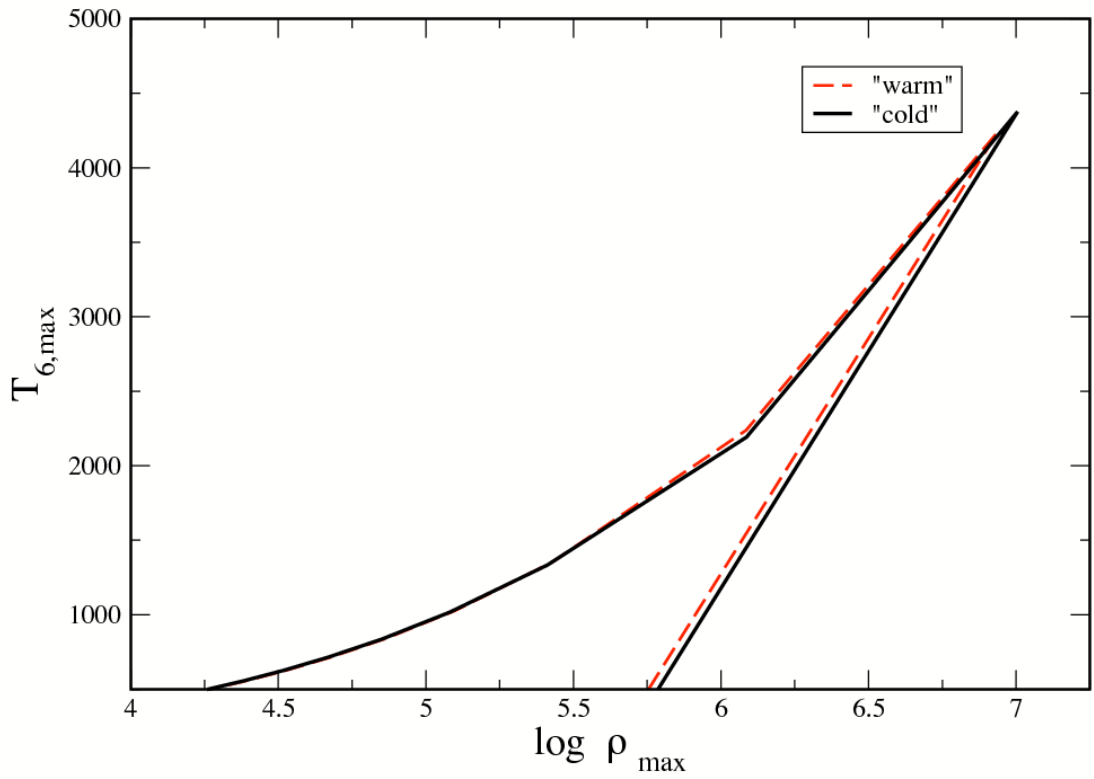}
\caption{Comparison of the average properties of the hottest 10\% of the SPH particles for an initially cold
and warm star (runs 8 and 9 in Table 1).}
\label{fig18}
\end{figure}

0.18 $M_\odot$ of Fe are synthesised when a typical 
0.6 \Msun C/O white dwarf approaches a 500 $M_\odot$ 
black hole with $\beta=5$. To explore how sensitive our models
are to the initial temperature, we have re-run this simulation, but
now with a hot white dwarf. The initial specific energy of a cold
($T= 5 \times 10^4$ K) equilibrium model was adjusted in a way that
the temperature varied linearly between $5 \times 10^7$ K in the centre
to $5 \times 10^4$ K at the stellar surface. In the subsequent relaxation
the star expanded slightly, since the degeneracy was lifted to a small 
extent. As a result of the reduced degeneracy, lower peak temperatures and  
densities are reached (see Figure~\ref{fig18}), reducing slightly the amount of  
nucleosynthesis which takes place. We consider the shown results
as representative, but the actual range of behaviour for given sets of masses
and orbits may be wider due to gradients in temperature and nuclear composition 
which have not been explored systematically here.
To summarize, in the most favorable cases, the nuclear energy release, is
comparable to that of typical Type Ia supernovae.

\begin{figure}
\epsscale{1.0} 
\plotone{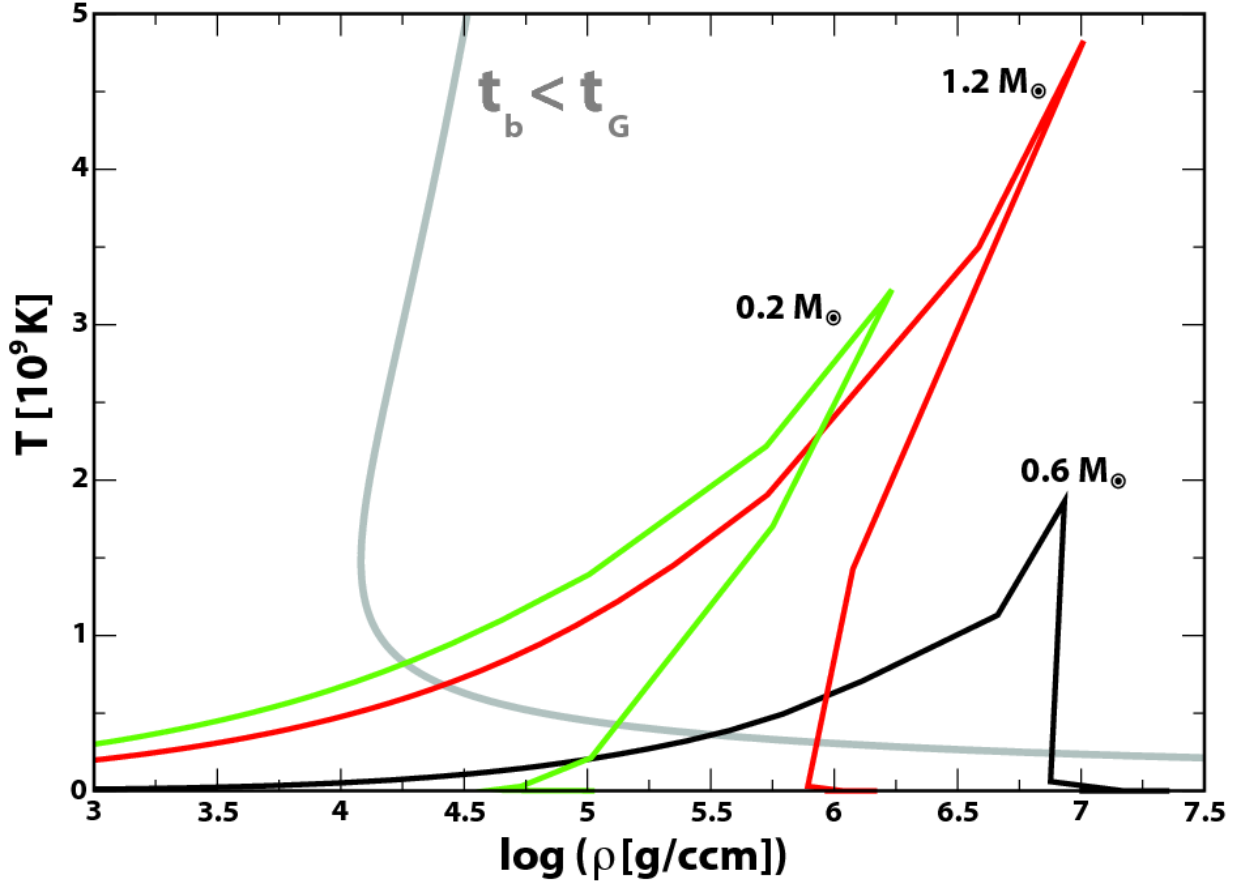}
\caption{The evolution of the compressed, and tidally disrupted white
  dwarfs of various masses in the $\rho-T$ plane: A 0.2 He $M_\odot$
  white dwarf approaching a $10^3 M_\odot$ black hole with $\beta=12$;
  a 0.6 C/0 $M_\odot$ white dwarf approaching a $10^4 M_\odot$ black
  hole with $\beta=1.5$ and a 1.2 C/0 $M_\odot$ white dwarf
  approaching a $10^3 M_\odot$ black hole with $\beta=1.5$. The
  hottest 10\% of the particles are identified and their average
  temperature (in units of $10^6$ K) is plotted as a function of their
  average density (in cgs units).  These trajectories always start
  cold and dense (right lower corner) and become hot and during the
  black hole flyby. For the 1.2 $M_\odot$ and 0.6 $M_\odot$ cases, the
  time scale on which the white dwarf can react is similar to the
  burning time scale and the star expands rapidly enough to quench
  explosive energy release. }
\label{fig19}
\end{figure}

In the tidal pinching process, explosive nucleosynthesis is likely to
proceed only for white dwarfs passing well within the tidal radius
(Figure~\ref{fig19}).  Explosive energy release, as calculated here, appears to be a
natural outcome for 0.6 $M_\odot$ (1.2 $M_\odot$) white dwarfs
approaching moderately massive black holes with $\beta \geq 5$ ($\beta
\geq 3$).

\section{Gravitational Waves}\label{gw}
\begin{figure}
\epsscale{1.0} 
\plotone{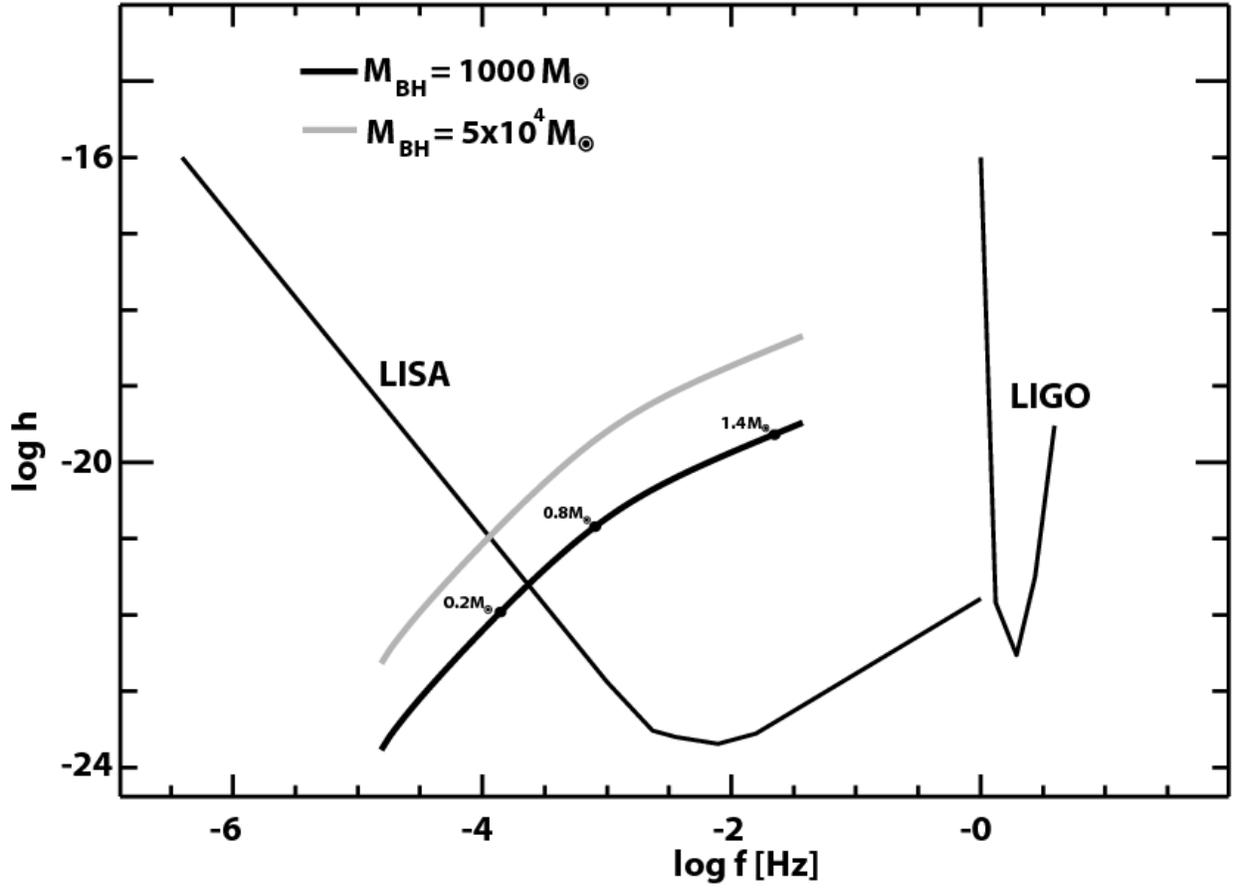}
\caption{Frequency and amplitude of gravitational radiation produced
  by the disruption of white dwarfs approaching an intermediate mass
  black hole, assuming $\beta=1$ and that the cluster hosting the
  black hole is at $D=$ 8kpc.}
\label{fig20}
\end{figure}

A white dwarf approaching the tidal radius will be disrupted in a
single flyby. The resulting stellar debris trail is not compact enough
to emit strong gravitational waves after leaving $R_{\rm p}$. The
detectable gravitational wave signal will therefore have a burst-like
behavior, roughly characterized by an amplitude $h$ and a duration $t
\sim 1/f$, where 
\be h \approx \frac{G}{c^4} \frac{\ddot{Q}}{D} \sim \frac{GM_{\rm wd} R_{\rm g}}{c^2 R_p D}
\sim 7 \times 10^{-20} \;\beta \paren{\frac{D}{\rm 10 kpc}}^{-1}
\paren{\frac{M_{\rm wd}}{0.6 M_\odot}}^{4/3} \paren{\frac{R_{\rm
      wd}}{10^9\;{\rm cm}}}^{-1} M_{\rm BH,3}^{2/3}, 
\ee
and 
\be f \sim
\paren{\frac{G M_{\rm BH}}{R_{\rm p}^3}}^{1/2} \sim 0.3\; \beta^{3/2}
\paren{\frac{M_{\rm wd}}{0.6 M_\odot}}^{1/2} \paren{\frac{R_{\rm
      wd}}{10^9\;{\rm cm}}}^{-3/2}\;{\rm Hz}.
\ee
LISA will be able to
detect gravitational waves of amplitude $h \sim 10^{-21}$ for burst
sources in the frequency range $f\sim 10^{-4}-10^{-1}$Hz
\citep{danzmann03}. Gravitational waves from white dwarf stellar
disruption could thus be detectable if $M_{\rm wd} \geq
0.4\;M_\odot$ and the source distance $D\leq 10$ kpc
(Figure~\ref{fig20}).

\begin{figure}
\epsscale{0.8} 
\plotone{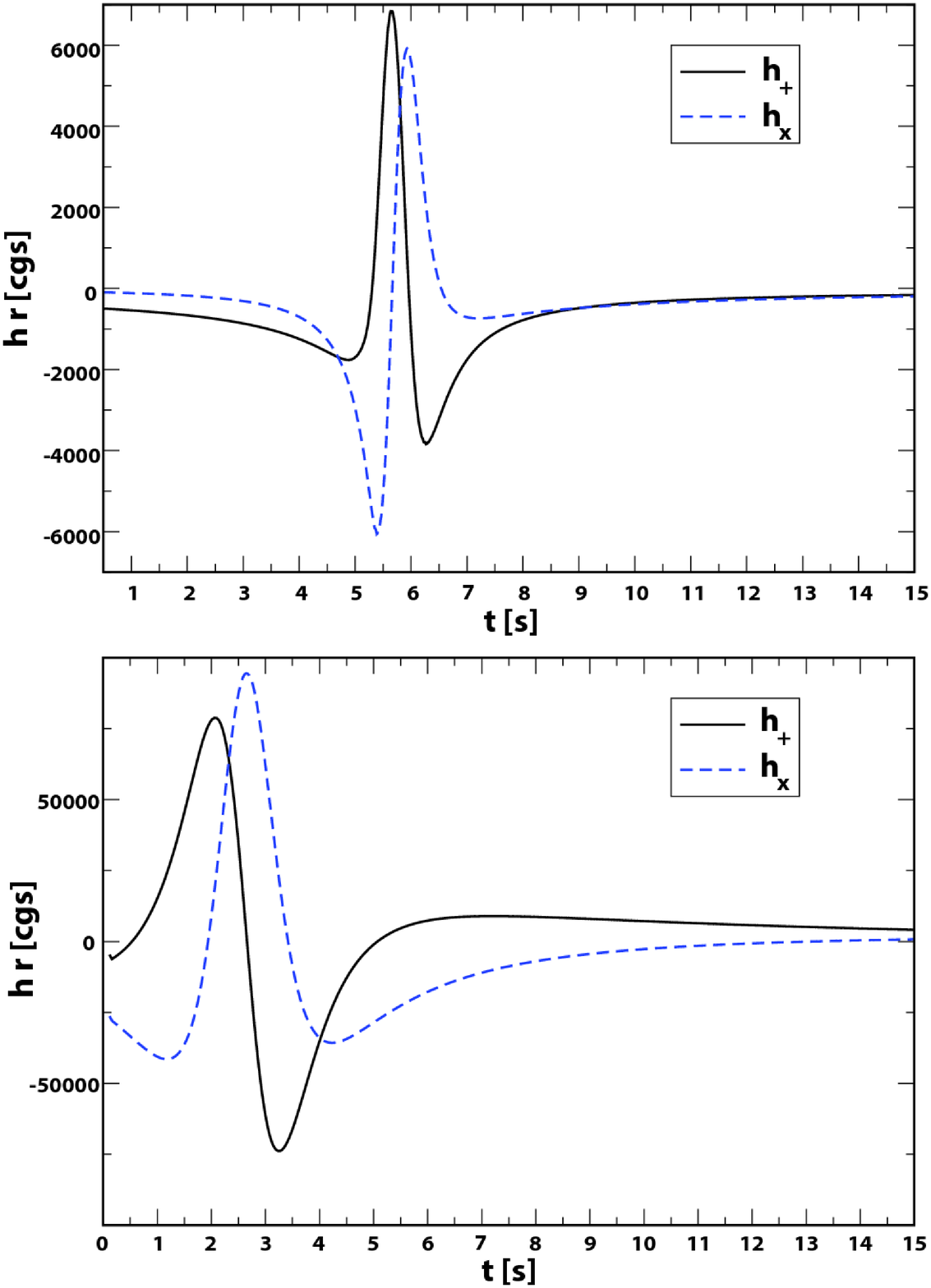}
\caption{Gravitational wave amplitudes. {\it Top panel:} gravitational
  radiation produced by the disruption of a 0.2 $M_\odot$ white dwarf
  approaching a $10^3 M_\odot$ black hole with $\beta=12$. {\it Bottom
    panel:} gravitational radiation produced by the disruption of a
  0.6 $M_\odot$ white dwarf approaching a $10^4 M_\odot$ black hole
  with $\beta=1.5$.}
\label{fig21}
\end{figure}

The gravitational wave amplitudes $h_{+}$ and $h_{\times}$ shown in
Figure~\ref{fig21} are calculated in the quadrupole approximation.
The reduced quadrupole moments can be written in terms of the
SPH-particle properties \citep{cm1993}
\be I_{jk}= \sum_i m_i (x_{ji} x_{ki} - \frac{1}{3} \delta_{jk}
r_i^2).  \ee The second time derivatives, $\ddot{I}_{jk}$, can then be
expressed in terms of the particle properties by simple, direct
differentiation.  The retarded gravitational wave amplitudes for a
distant observer along the z -axis at distance $D$ are given by \be D
\, h_{+}= \frac{G}{c^{4}}(\ddot{I}_{xx}-\ddot{I}_{yy}) \ee and \be D
\, h_{\times}= 2 \frac{G}{c^{4}}\ddot{I}_{xy}.  \ee Both duration and
amplitudes of the gravitational wave bursts that are produced by the
disruption calculations (Figure~\ref{fig21}) are in agreement with the
simple estimates given above.

\section{Discussion}\label{dis}
This paper presents a computational investigation of the mechanical
and nuclear evolution of white dwarfs passing well within the 
tidal radius of a moderately massive black hole. A comprehensive 
top to bottom approach is adopted: we follow the tidal disruption and
compression leading up to the ignition of the white dwarf; the complex
propagation of the nuclear energy release through the star; the
resultant gravitational wave signal and the subsequent accretion of
the bound debris.

This paper has outlined several potentially observable effects. The
detection of a peculiar, underluminous thermonuclear explosion
\citep{rrh2008,k2008} accompanied by a thermal transient signal of
predominantly soft X-rays with a peak luminosity $L\sim L_{\rm
  Edd}=10^{41}M_{BH,3}$ erg/s, fading within a year would, if detected,
be a compelling testimony for the existence of a new mechanism by
which white dwarf ignition can be achieved.  Although the
thermodynamical and nuclear energy of the star is sensitive to the
initial composition, we found that thermonuclear
ignition is a natural outcome for
white dwarfs of all masses passing well within the tidal radius, with
lighter stars requiring deeper penetration into the tidal radius
due to their lower densities. For simplicity, we have instantiated 
our initial carbon oxygen white dwarf models  (M$_{\rm WD} \ge 0.6$ 
\msun)  as homogeneously mixed stars with a 50\% mass fraction of 
each nucleus. While such internal chemical profiles are likely 
accurately realized in nature in very massive white dwarfs 
($\sim 1$ \msun) \citep{mazzitelli86}, for lower masses the gravothermal 
adjustment of the interior during the cooling phase produces 
oxygen-enhanced stellar cores surrounded by very carbon-rich mantles 
($X_C \sim 0.8$). The exact radial distribution depends on the exact 
value of $^{12}C(\alpha,\gamma)^{16}O$ rate and the details of how 
convection proceeds \citep{mazzitelli86,salaris97,straniero03}, but this 
general stratification tendency is  well-established. Thus, the disruption 
of a standard 0.6 \Msun white dwarf should produce a highly carbon-enriched 
remnant atmosphere. Maybe, the recently detected carbon-rich transient 
SCP 06F6 accompanied by an X-ray signal \citep{gaensicke08} is already 
the first example for this class of object.

The gravitational forces from the central black hole are currently
calculated using a Paczy\'nski-Wiita pseudo potential but our goal is
to incorporate the effects of general relativity, as white dwarfs
passing well within the tidal radius cannot be disrupted without
entering the strong relativistic regime. The form of the black hole
(Schwarzschild or Kerr?) then has an important quantitative effect, as
does (for a rotating Kerr hole) the orientation of the stellar orbit
relative to the  black hole spin axis. The orbits are then not ellipses, but
may turn through 2$\pi$ or even more \citep{lu1989}. A fluid element,
which in the case of an elliptical orbit would cross the orbital plane
just once, may then have two or more traversals. This opens up the
possibility of multiple shocks.

For a white dwarf which does not pass close enough to the hole to
release much energy on a dynamical timescale, adiabatic cooling would
severely reduce the internal radiative content before the debris
became translucent (just as a supernova would be optically
inconspicuous in the absence of continuing energy injection in the
months after the explosion). There would be no transient until, as
discussed above, the bound debris fell back onto the hole after
$t_{\rm fb}$.  The integrated output from this flare could, in
principle, amount to a few per cent of the white dwarf's rest mass,
but would probably be significantly less, because most of the debris
would be fed to the black hole far more rapidly than it could be
accepted if the radiative efficiency were high; much would then be
swallowed inefficiently or most likely escape in radiatively-driven
directed outflow: its ram pressure and subsequent heating could
inhibit the steady accretion that would otherwise be inevitable in any
galaxy or globular cluster harboring a moderately massive black hole.

\acknowledgments We thank Holger Baumgardt, Peter Goldreich, Jim Gunn,
Piet Hut, Dan Kasen, Bronson Messer and Martin Rees for very useful
discussions. E. R. acknowledges support from the DOE Program for
Scientific Discovery through Advanced Computing (SciDAC;
DE-FC02-01ER41176). The simulations presented in this paper were 
performed on the JUMP computer of the H\"ochstleistungsrechenzentrum J\"ulich.
Oak Ridge National Laboratory is managed by UT-Battelle, LLC, for the  
U.S. Department of Energy under contract DE-AC05-00OR22725.

\newpage
\begin{table}
\caption{Summary of the performed runs. $M_{\rm wd}$ and  $M_{\rm BH}$ are the masses of the white dwarf and the black hole, respectively, $\beta$ is the ratio of tidal radius and pericentre distance. The type of gravity is indicated by N (Newtonian) or PW (Paczy\'nski-Wiita). Column six states the number of SPH particles used in the simulation, $E_{\rm burn}$ is the energy generated in burning processes. To put this number into context we briefly state the gravitational binding energies (in ergs) of the different white dwarfs: $\log(E_{{\rm bin,}0.2 M_\odot})= 49.13$, $\log(E_{{\rm bin,}0.6 M_\odot})= 50.12$, $\log(E_{{\rm bin,}1.2 M_\odot})= 50.98$. ``Fe'' labels the mass in iron-group elements, ``expl.'' in the comment column indicates that the produced nuclear energy exceeds the WD gravitational binding energy.}
\begin{tabular}{llrrlrlll}
\hline
run & $M_{\rm wd}$ & $M_{\rm BH}$ & $\beta$ & grav & SPH part. & log($E_{\rm burn}$) & ``Fe'' [\msun]&comments\\
\hline
1   & 0.2         & 1000       & 12      & N    & 4034050   & 50.46   &   0.025   & expl.      \\
2   & 0.2         & 1000       & 12      & PW   & 4034050   & 50.44   &   0.034   & expl.     \\
3   & 0.2         & 1000       & 12      & PW   &  200452   & 50.44   &           & $\Gamma=5/3$-polytrope   \\
4   & 0.2         &  100       &  5      & PW   &  100027   & 49.57   &$<10^{-10}$ & explore BH influence, expl.  \\
5   & 0.2         &  500       &  5      & PW   &  100027   & 49.64   &$<10^{-10}$ & explore BH influence, expl. \\
6   & 0.2         & 1000       &  5      & PW   &  100027   & 49.76   &$<10^{-10}$ & explore BH influence, expl. \\
7   & 0.2         & 5000       &  5      & PW   &  100027   & 49.93   &$<10^{-10}$ & explore BH influence, expl. \\
8   & 0.6         &  500       &  5      & N    &  502479   & 50.68   & 0.18      & expl.       \\
9   & 0.6         &  500       &  5      & N    &  502479   & 50.62   & 0.13      & hot, initial WD \\
10  & 0.6         & 1000       &  0.9    & N    & 1006446   &  0.00   & 0.        & no nuclear burning       \\
11  & 0.6         & 1000       &  5      & PW   &  502479   & 50.43   & $3\times 10^{-4}$&  \\
12  & 0.6         &10000       &  1.5    & PW   &  502479   & 45.07   &$<10^{-10}$ &        \\
13  & 1.2         &  100       &  3.5    & N    &  100027   & 51.01   &   0.58    & expl.      \\
14  & 1.2         &  500       &  2.6    & PW   &  502479   & 51.16   &   0.66    & expl.       \\
15  & 1.2         & 1000       &  1.5    & PW   &  502479   & 49.63   & $0.014 $  & \\
16  & 1.2         & 1000       &  3.0    & N    &  502479   & 51.10   & 0.63      & expl.       \\
\end{tabular}
\label{tab:runs}
\end{table}

\clearpage

\clearpage

\newpage

\newpage

\newpage

\newpage

\newpage

\newpage

\newpage

\newpage

\newpage

\newpage

\newpage

\newpage

\newpage

\newpage

\newpage

\newpage

\newpage

\clearpage

\clearpage

\end{document}